\documentclass[10pt]{article}

\usepackage{amsmath}

\usepackage{amssymb}

\usepackage{graphicx}

\usepackage{cite}

\usepackage{color} 

\usepackage[labelfont=bf,labelsep=period,justification=raggedright]{caption}

\makeatletter
\renewcommand{\@biblabel}[1]{\quad#1.}
\makeatother

\date{}

\begin{document}

\begin{flushleft}
{\Large
\textbf{Bilayer elasticity at the nanoscale: the need for new terms}
}
\\
Anne-Florence Bitbol$^{1, \dag}$, 
Doru Constantin$^{2}$, 
Jean-Baptiste Fournier$^{1,\ast}$
\\
\bf{1} Laboratoire Mati\`ere et Syst\`emes Complexes (MSC), Universit\'e Paris Diderot, Paris 7, Sorbonne Paris Cit\'e, CNRS UMR 7057, Paris, France
\\
\bf{2} Laboratoire de Physique des Solides, Universit\'e Paris-Sud, Paris 11, CNRS UMR 8502, Orsay, France
\\
$\ast$ E-mail: jean-baptiste.fournier@univ-paris-diderot.fr
\\
$\dag$ Current address: Lewis-Sigler Institute for Integrative Genomics, Princeton University, Princeton, New Jersey, United States of America
\end{flushleft}

\section*{Abstract}

Continuum elastic models that account for membrane thickness variations are especially useful in the description of nanoscale deformations due to the presence of membrane proteins with hydrophobic mismatch. We show that terms involving the gradient and the Laplacian of the area per lipid are significant and must be retained in the effective Hamiltonian of the membrane. We reanalyze recent numerical data, as well as experimental data on gramicidin channels, in light of our model. This analysis yields consistent results for the term stemming from the gradient of the area per molecule. The order of magnitude we find for the associated amplitude, namely 13--60 mN/m, is in good agreement with the 25 mN/m contribution of the interfacial tension between water and the hydrophobic part of the membrane. The presence of this term explains a systematic variation in previously published numerical data. 

\section*{Introduction}

As basic constituents of cell membranes, lipid bilayers \cite{Mouritsen_book} play an important role in biological processes, not as a passive background, but rather as a medium that responds to and influences, albeit in a subtle way, the behavior of other membrane components, such as membrane proteins \cite{Sackmann84}. The coupling between the lipid bilayer and guest molecules does not occur by the formation of chemical bonds, but rather by a deformation of the membrane in its entirety. To describe it, one must resort to concepts developed in soft matter physics for the understanding of self-assembled systems.

At length scales much larger than their thickness, the elasticity of lipid bilayers is well described by the Helfrich model~\cite{Helfrich73}. However, nanometer-sized inclusions, such as membrane proteins, deform the membrane over smaller length scales. In particular, some transmembrane proteins have a hydrophobic part with a thickness slightly different from that of the hydrophobic part of the membrane. Due to this hydrophobic mismatch, the hydrophobic core of the membrane locally deforms~\cite{Owicki78, Owicki79, Mouritsen84}. As this deformation affects the thickness of the membrane, and as its characteristic amplitude and decay length are both of a few nanometers~\cite{Huang86}, it cannot be described using the Helfrich model. In fact, since the range of such deformations is of the same order as membrane thickness, one can wonder to what extent continuum elastic models in general still apply, and what level of complexity is required for an accurate description. In particular, which terms must be retained in a deformation expansion of the effective Hamiltonian?

Experimental data is available for the gramicidin channel~\cite{Kelkar07}, a transmembrane protein formed by two protein monomers. The channel being large enough for the passage of monovalent cations, conductivity measurements~\cite{OConnell90} can detect its formation and lifetime, which are directly influenced by membrane properties. The gramicidin channel can therefore act as a local probe for bilayer elasticity on sub-nanometer scales (see, e.g., Ref.~\cite{Lundbaek10}). Motivated by this opportunity, sustained theoretical investigations have been conducted in order to construct a model describing membrane thickness deformations~\cite{Huang86,Helfrich90,Dan93,ArandaEspinoza96}. Recently, detailed numerical simulations have been performed, giving access both to the material constants involved in elastic models and to the membrane shape close to a mismatched protein~\cite{Brannigan06, Brannigan07, West09}. This numerical data provides a good test for theoretical models.

In this article, we put forward a modification to the models describing membrane thickness deformations. We argue that contributions involving the gradient (and the Laplacian) of the area per lipid should be accounted for in the effective Hamiltonian per lipid from which the effective Hamiltonian of the bilayer is constructed, following the approach of Refs.~\cite{Dan93, ArandaEspinoza96}. We show that these new terms cannot be neglected, as they contribute to important terms in the bilayer effective Hamiltonian. We discuss the differences between our model and the existing ones. We compare the predictions of our model with numerical data giving the profile of membrane thickness close to a mismatched protein~\cite{Brannigan06, Brannigan07, West09}, and with experimental data on gramicidin lifetime~\cite{Elliott83} and formation rate~\cite{Goulian98}.

\section*{Results: Membrane model}

We consider a bilayer membrane constituted of two identical monolayers, labeled by $+$ and $-$, in contact with a reservoir of lipids with chemical potential $\mu$. We write the effective Hamiltonian per molecule in monolayer $\pm$ as
\begin{align}
f_m^\pm&=\frac{1}{2}f''_0(\Sigma^\pm-\Sigma_0)^2\pm f_1\,H^\pm \pm f'_1(\Sigma^\pm-\Sigma_0)H^\pm+f_2\,(H^\pm)^2\nonumber\\
&+f_K\,K^\pm +\alpha\,(\mathbf{\nabla}\Sigma^\pm)^2+\beta\,\nabla^2\Sigma^\pm+\zeta\,(\nabla^2\Sigma^\pm)^2-\mu\,,
\label{fmolec}
\end{align}
where $\Sigma^\pm$ is the area per lipid, while $H^\pm$ is the local mean curvature of the monolayer, and $K^\pm$ is its local Gaussian curvature (denoting by $c_1^\pm$ and $c_2^\pm$ the local principal curvatures~\cite{Safran} of the monolayer, we have $H^\pm=(c_1^\pm+c_2^\pm)/2$ and $K^\pm=c_1^\pm c_2^\pm$). All these quantities are defined on the hydrophilic-hydrophobic interface of each monolayer. Eq.~\ref{fmolec} corresponds to an expansion of $f_m^\pm$ for small deformations around the equilibrium state where the membrane is flat and where each lipid has its equilibrium area $\Sigma_0$. Any constant term in the free energy per lipid is included in a redefinition of the chemical potential $\mu$. From now on, we will consider small deformations of an infinite flat membrane and we will work in the Monge gauge, so $H^\pm\simeq\nabla^2 h^\pm/2$ and $K^\pm\simeq\partial^2_x h^\pm \partial^2_y h^\pm-(\partial_x\partial_y h^\pm)^2=\det(\partial_i\partial_j h^\pm)$, where $z=h^\pm(x,y)$ represents the height of the hydrophilic-hydrophobic interface of each monolayer with respect to a reference plane $(x,y)$. The upper monolayer is labeled by~$+$ and the lower one by~$-$.
Many constants involved in Eq.~\ref{fmolec} can be related to the constitutive constants of a monolayer: $f''_0\Sigma_0=K_a/2$ is the compressibility modulus of the monolayer, $f_2/(2\Sigma_0)=\kappa_0/2$ is its bending rigidity, $f_K/\Sigma_0=\bar\kappa/2$ is its Gaussian bending rigidity, $f_1/f_2=c_0$ is its spontaneous (total) curvature, and $f'_1/f_2=c'_0$ is the modification of the spontaneous (total) curvature due to area variations (see Methods, Sec.~\ref{Deriv}).

In the case where $\alpha=\beta=\zeta=0$, Eq.~\ref{fmolec} is equivalent to the model of Ref.~\cite{Safran}, which is the basis of that developed in Refs.~\cite{Dan93, ArandaEspinoza96, Brannigan06, Brannigan07}. To our knowledge, existing membrane models including the area per lipid (or, equivalently, the two-dimensional lipid density) do not explicitly feature terms in the gradient, or Laplacian, of this variable~\cite{Bitbol11_stress}. The possibility of an independent term proportional to the squared thickness gradient was however suggested on symmetry grounds in Ref.~\cite{Fournier99}, while pointing that it could arise from the specific cost of modulating the area per lipid (see note (18) in Ref.~\cite{Fournier99}). In the present work, we show that the terms in $\alpha$, $\beta$ and $\zeta$ cannot be neglected with respect to others. We focus on the influence of $\alpha$, for which we provide a physical interpretation, and we will set $\beta=\zeta=0$ in the body of this article in order to simplify our discussion and to avoid adding unknown parameters. However, the derivation of the membrane effective Hamiltonian is presented in Secs.~\ref{Deriv}-\ref{elim_deltat} of our Methods part, in the general case where $\alpha$, $\beta$ and $\zeta$ are all included. 

\begin{figure}[h t b]
\centering
\includegraphics[width=0.9\textwidth]{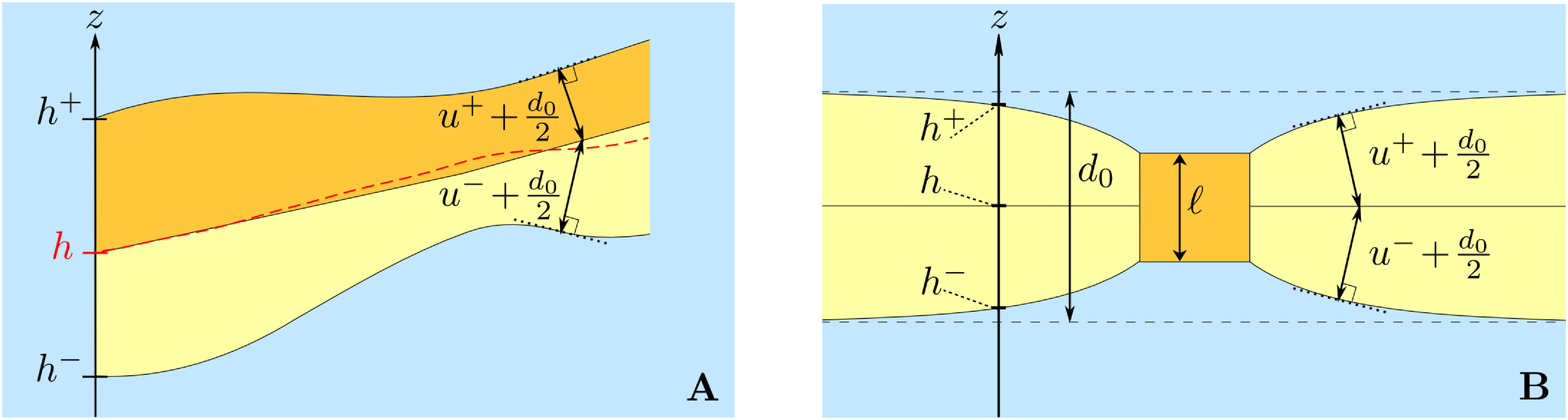}
\caption[]{\textbf{Definitions.} \textbf{A)} Cut of a bilayer membrane. The solid black lines mark the boundaries of the hydrophobic part of the membrane, and the exterior, which is shaded in blue, corresponds to the hydrophilic lipid heads and the water surrounding the membrane. The hydrophobic thickness, defined along the normal to the hydrophobic-hydrophilic interface, of the upper (resp. lower) monolayer, shaded in orange (resp. yellow), is $u^++d_0/2$ (resp. $u^-+d_0/2$). The height of monolayer $\pm$ along $z$ is denoted by $h^\pm$. The average membrane shape, $h=(h^++h^-)/2$, is represented as a red dashed line. \textbf{B)} Cut of a bilayer membrane (with hydrophobic part shaded in yellow) containing a protein with a hydrophobic mismatch (orange square). The equilibrium hydrophobic thickness of the bilayer is $d_0$, while the hydrophobic thickness of the protein is $\ell$. The average shape of the membrane is flat, and the thickness deformations of the two monolayers are identical ($u^+=u^-=u/2$). Hence, the average shape $h$ is constant, and confounded with the midlayer of the membrane. Although $u^\pm$ is defined along the normal to the monolayer hydrophilic-hydrophobic interface, the boundary condition at the inclusion edge, i.e., in $r=r_0$, simply reads $u(r_0)=u_0=\ell-d_0$ to first order (see main text, Section entitled ``Deformation profiles close to a mismatched protein'').
\label{Dessin_defs}}
\end{figure}

The effective Hamiltonian of a bilayer membrane patch with projected area $A_p$ at chemical potential $\mu$ can be derived from Eq.~\ref{fmolec}. For this, the effective Hamiltonians per unit projected area of the two monolayers are summed, taking into account the constraint that there is no space between the two monolayers of the bilayer, and assuming that the hydrophobic chains of the lipids are incompressible. This derivation is carried out in Sec.~\ref{Deriv} of our Methods part. It results in an effective Hamiltonian of the bilayer membrane that depends on three variables: the average shape $h=(h^++h^-)/2$ of the bilayer, the sum $u$ of the excess hydrophobic thicknesses of the two monolayers, each being measured along the normal to the monolayer hydrophilic-hydrophobic interface (see Fig.~\ref{Dessin_defs} and Eqs.~\ref{def_var}-\ref{jbar}), and the difference $\delta$ between the monolayer excess hydrophobic thicknesses. (The excess hydrophobic thickness of a monolayer is defined as the hydrophobic thicknesses of this monolayer minus its equilibrium value.)

In the present work, we are not interested in the degree of freedom $\delta$, which is not excited in the equilibrium shape of a membrane containing up-down symmetric mismatched proteins (see see Fig.~\ref{Dessin_defs}B). Hence, in Sec.~\ref{elim_deltat} of our Methods part, we integrate $\delta$ out, which amounts to minimizing $f$ with respect to $\delta$ since our theory is Gaussian. 
The resulting effective Hamiltonian, which involves $h$ and $u$, is given by Eq.~\ref{moregenbis} in Sec.~\ref{elim_deltat} of our Methods part. In this effective Hamiltonian, the variables $h$ and $u$ are decoupled, and the part depending on $h$ corresponds to the Helfrich Hamiltonian~\cite{Helfrich73}. Hence, our model gives back the Helfrich Hamiltonian if the state of the membrane is described only by its average shape $h$ (see Methods, Sec.~\ref{special_cases}). 

Here, we focus on variations of the membrane thickness, i.e., on the variable $u$. We thus restrict ourselves to the case where the average shape $h$ of the membrane is flat (see Fig.~\ref{Dessin_defs}B). In this case, we obtain, from Eq.~\ref{moregenbis}:
\begin{align}
f&=\frac{\sigma}{d_0}\,u+\frac{K_a}{2\,d_0^2}\,u^2+\frac{K'_a}{2}\,(\mathbf{\nabla}u)^2+\frac{K''_a}{2}\,(\nabla^2 u)^2\nonumber\\
&+A_1\,\nabla^2 u+A_2\,\mathbf{\nabla}\cdot(u\mathbf{\nabla}u) +\frac{\bar\kappa}{4}\,\det(\partial_i\partial_j u)\,.
\label{jbarc2b}
\end{align}
In the case where $\beta=\zeta=0$, on which the body of this article focuses, the various constants introduced in Eq.~\ref{jbarc2b} read:
\begin{align}
\sigma&=-\frac{2\mu}{\Sigma_0}\,,\\
K'_a&=-\frac{\kappa_0}{d_0}(c_0-c'_0\Sigma_0)+k'_a+\frac{\sigma}{4}\,,\label{Kpa}\\
K''_a&=\frac{\kappa_0}{4}\,,\label{Ksa}\\
A_1&=\frac{\kappa_0 \,c_0}{2}\,,\\
A_2&=\frac{\kappa_0}{2\,d_0}(c_0-c'_0\Sigma_0)\,.\label{A2}
\end{align}
In these equations, $d_0$ denotes the equilibrium hydrophobic thickness of the bilayer membrane, $\sigma$ plays the part of an externally applied tension (see Methods, Sec.~\ref{Tension}), $K_a$ is the compressibility modulus of the membrane, $\bar\kappa$ is its Gaussian bending rigidity, $\kappa_0$ is the bending rigidity of a symmetric membrane such that $\delta=0$, $c_0$ is the spontaneous (total) curvature of a monolayer, and $c'_0$ is the modification of this spontaneous curvature due to area variations. In addition, we have introduced $k'_a=4\,\alpha\,\Sigma_0/d_0^2$, which has the dimension of a surface tension, like $K_a$. Note that the last three terms in Eq.~\ref{jbarc2b} are boundary terms. 

In Sec.~\ref{elim_deltat} of our Methods part, the expressions of $K'_a$, $K''_a$, $A_1$ and $A_2$ are provided in the more general case where $\beta$ and $\zeta$ are included.

We wish to describe a membrane with an equilibrium state that corresponds to a homogeneous thickness. A linear stability analysis (presented in Sec.~\ref{Stab} of our Methods part) shows that the flat shape is stable if $K_a>0$, $K''_a>0$, and
\begin{align}
K'_a>-2\frac{\sqrt{K_a K''_a}}{d_0}\,.
\label{stab}
\end{align}

\section*{Discussion}
\subsection*{Comparison with existing models}

Our model Eq.~\ref{jbarc2b} has a form similar to that of the models developed in Refs.~\cite{Dan93, ArandaEspinoza96, Brannigan06, Brannigan07}. However, it differs from these previous models on several points. First, our definition of $u$ is slightly different. Second, we have included the effect of an applied tension $\sigma$. Finally, the various constants in Eq.~\ref{jbarc2b} have different interpretations, and thus different values, from the ones in the existing models. Let us discuss these points in more detail.

\subsubsection*{On the definition of $u$}
In the present work, the variable $u$, which is the relevant one to study membrane thickness deformations, is defined as the sum of the excess hydrophobic thicknesses of the two monolayers, each being measured along the normal to the monolayer hydrophilic-hydrophobic interface (see Eqs.~\ref{def_var}-\ref{jbar} in the Methods section). This definition of $u$ has the advantage of being independent of deformations of the average membrane shape $h$. 

The excess thickness variable used in Refs.~\cite{Huang86, Goulian98, Nielsen98, Dan93, ArandaEspinoza96, Brannigan06, Brannigan07, Watson11} reads in our notations:
\begin{equation}
\bar u=\frac{h^+-h^--d_0}{2}\,. \label{uprime}
\end{equation}
Using Eqs.~\ref{uprime} and~\ref{ctrg}, and working to second order, we obtain
\begin{equation}
2\,\bar u=u-\frac{d_0}{2}\left[(\mathbf{\nabla}h)^2+\frac{(\mathbf{\nabla}u)^2}{4}\right]\,, \label{ladiff}
\end{equation}
which shows that there is a second-order difference between $2\,\bar u$ and our variable $u$. Consequently, the difference between the definition used in the previous works and ours regards only the term linear in $u$, i.e., the tension term, which was not included in these works. At zero applied tension, the two definitions are equivalent, i.e., it is equivalent to use $u$ or $2\,\bar u$. Our definition of $u$ is the right one for rigorously taking tension into account, because it is independent of deformations of the average membrane shape $h$: the energy stored in the variable $u$ only comes from thickness variations. (The variable $\bar u$ of Refs.~\cite{Huang86, Goulian98, Nielsen98, Dan93, ArandaEspinoza96, Brannigan06, Brannigan07, Watson11} corresponds to the difference between the bilayer hydrophobic thickness \emph{projected along $z$} and the \emph{non-projected} equilibrium hydrophobic bilayer thickness (see Eq.~\ref{uprime}), so it is not independent of $h$. The second-order difference between $2\,\bar u$ and $u$, which is shown in Eq.~\ref{ladiff}, arises from this difference in projection between actual thicknesses and equilibrium thicknesses within the definition of $\bar u$.)

\subsubsection*{On tension}
First of all, existing models~\cite{Huang86, Goulian98, Nielsen98, Dan93, ArandaEspinoza96, Brannigan06, Brannigan07} were constructed at zero applied tension, which means $\sigma=0$ in Eq.~\ref{jbarc2b}. To our knowledge, our work is the first where the coefficient of the term linear in $u$ is explicitly related to the applied tension (see Methods, Sec.~\ref{Tension}) and to the tension of the Helfrich model (see Methods, Sec.~\ref{special_cases}). 

In Ref.~\cite{Goulian98}, the effect of applied tension is taken into account, in so far as it changes the equilibrium membrane thickness of a homogeneous membrane, but without being fully implemented in the elastic model. Our more complete description gives back this effect on membrane thickness, when it is applied to the particular case of a homogeneous membrane (see Methods, Sec.~\ref{Tension}).

\subsubsection*{On the constant $K'_a$}
In our model, the constant $K'_a$ features three contributions with different origins (see Eq.~\ref{Kpa}). 

The first contribution arises from the spontaneous curvature of a monolayer and from its variation with the area per lipid. More precisely, the term 
\begin{equation}
\frac{\kappa_0}{2\,d_0}(c_0-c'_0\Sigma_0)\,u\nabla^2 u=\frac{\kappa_0}{2\,d_0}(c_0-c'_0\Sigma_0)\,\left[\mathbf{\nabla}\cdot(u\mathbf{\nabla}u)-(\mathbf{\nabla}u)^2\right]
\end{equation}
appears when one constructs the membrane model starting from a monolayer Hamiltonian density such as Eq.~\ref{fmolec}. This term was first introduced in Ref.~\cite{Dan93}, and it was then included in Refs.~\cite{ArandaEspinoza96, Brannigan06}.

\begin{figure}[h t b]
\centering
\includegraphics[width=0.15\textwidth]{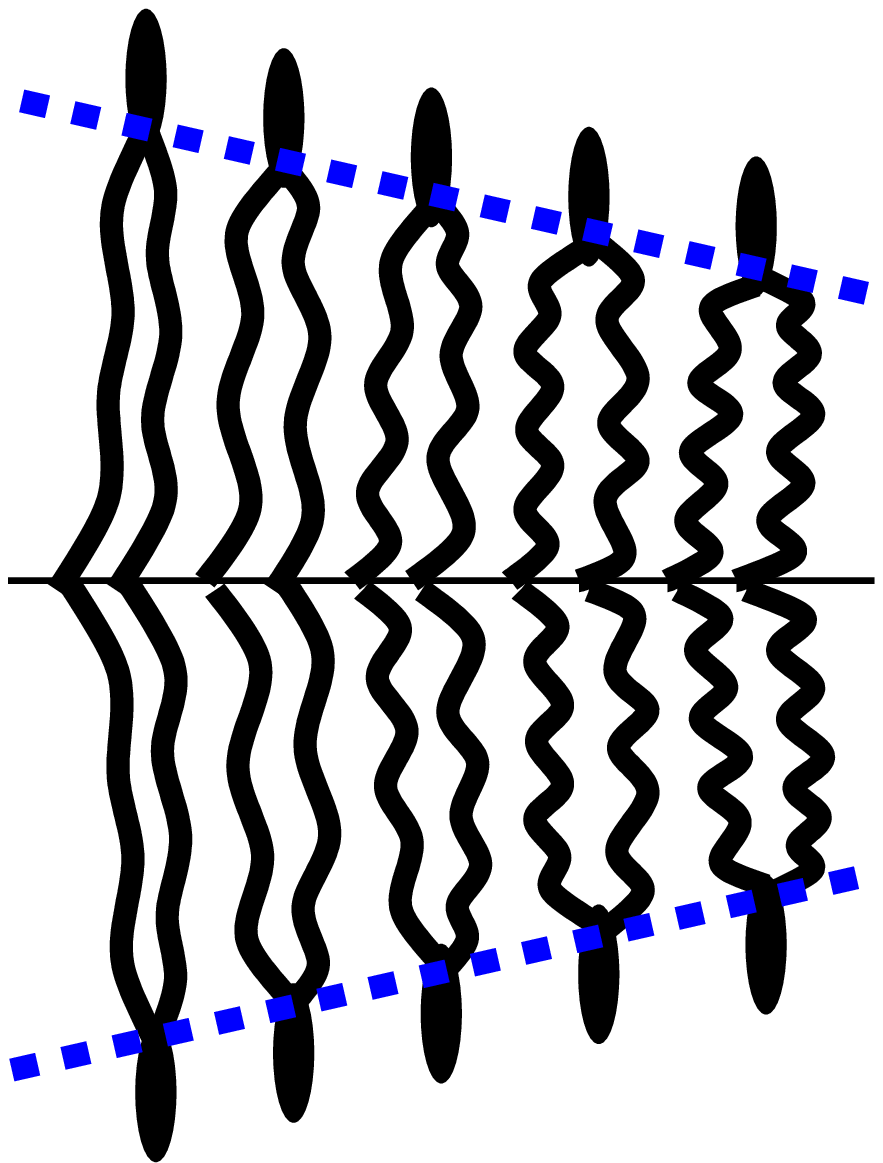}
\caption[]{\textbf{Thickness gradient.} Cut of a bilayer membrane with a symmetric thickness gradient. The dashed blue lines correspond to the hydrocarbon-water interfaces.\label{Figep}}
\end{figure}

The second contribution, $k'_a$, arises from $\alpha$, i.e., from the term in $(\mathbf{\nabla}\Sigma)^2$ introduced in Eq.~\ref{fmolec}. This term was not included in Refs.~\cite{Dan93, ArandaEspinoza96, Brannigan06}, which started from a second-order expansion of the effective Hamiltonian per lipid molecule involving only the curvature and the area per lipid. However, a gradient of area per lipid (or, equivalently, of the thickness) in a monolayer has an energetic cost of its own. Indeed, a greater part of the hydrophobic chains is in contact with water when a thickness gradient is present (see Fig.~\ref{Figep}). The associated energetic cost is given by the interfacial tension $\gamma$ of the hydrocarbon-water interface, which is of order 40--50~mN/m~\cite{Israelachvili, Sharp91}. Such a term is often accounted for in microscopic membrane models (see, e.g., Ref.~\cite{May99}). In the case of a symmetric membrane ($u^+=u^-=u/2$) with flat average shape, the surface of the hydrocarbon-water interface is increased by a factor $[1+(\mathbf{\nabla}u)^2/8]$ for each monolayer (see Fig.~\ref{Figep}). Thus, to second order, the associated energetic cost per unit projected area is $\gamma (\mathbf{\nabla}u)^2/4$. Note that other physical effects, e.g., the elasticity of the chains, may yield contributions to the term in $(\mathbf{\nabla}\Sigma)^2$. However, if we restrict to the simple term arising from interfacial tension, we obtain 
\begin{equation}
 k'_a=\frac{\gamma}{2}\approx25\,\mathrm{mN/m}\,. 
\label{kpa_g}
\end{equation}

Finally, the third contribution, $\sigma/4$, arises from the (macroscopic) externally applied tension. The tension of a vesicle can rise only up to a few mN/m before it bursts (see, e.g., Ref.~\cite{Goulian98}). Hence, according to our estimate of $k'_a$ in Eq.~\ref{kpa_g}, we expect $\sigma/4\ll k'_a$.

In the seminal article Ref.~\cite{Huang86}, where the membrane model was constructed by analogy with liquid crystals, a term in $(\mathbf{\nabla}u)^2$, interpreted as arising from tension, was included in the effective Hamiltonian. However, its effect was neglected on the grounds that the value of its prefactor made it negligible with respect to the other terms. The value of this prefactor was taken to be that of the tension of a monolayer on the surface of a Plateau border~\cite{Hladky82}. The model introduced in Ref.~\cite{Huang86} was further developed and analyzed in Refs.~\cite{Goulian98, Nielsen98}, where the same argument was used to neglect the term in $(\mathbf{\nabla}u)^2$. 

However, our construction of the membrane effective Hamiltonian shows that the microscopic tension involved through $k'_a$ arises from local variations in the area per lipid. This stands in contrast with the case of the Plateau border, where whole molecules can move along the surface and exchange with the bulk, yielding a smaller value of the tension. Ref.~\cite{Hladky82} stresses that the measured tension of a Plateau border is valid for long-wavelength fluctuations, but that it is largely underestimated for short-wavelength fluctuations (less than 10~nm) which involve significant changes in area per molecule. 

Including the tension of the hydrocarbon-water interface instead of that of the Plateau border is a significant change, given that the former is of order 40--50~mN/m~\cite{Israelachvili, Sharp91}, while the latter is of order 1.5--3~mN/m~\cite{Hladky82, Huang86, Goulian98, Nielsen98}. In Refs.~\cite{Goulian98, Nielsen98}, it is shown that the effect of the term in $(\mathbf{\nabla}u)^2$ is negligible if
\begin{equation}
K'_a\ll\frac{\sqrt{K_a K''_a}}{d_0}\,,
\label{negl}
\end{equation}
where we have used our own notations of the prefactors of the terms in $(\mathbf{\nabla}u)^2$, $u^2$ and $(\nabla^2u)^2$. In the case of DOPC, taking $K''_a=\kappa/4$ and using the values of the membrane constants~\cite{Rawicz00}, this condition becomes $K'_a\ll 28\,\mathrm{mN/m}$. While this is well verified if $K'_a$ corresponds to the tension of the Plateau border, it is no longer valid within our model.

Our model is the first that includes all contributions to $K'_a$, in particular the one arising from interfacial tension. Besides, in Sec.~\ref{elim_deltat} of our Methods part, we show that $\beta$ is also involved in $K'_a$, which emphasizes the complexity of constructing a continuum model to describe membrane elasticity at the nanoscale: many terms involved in the expansion of the effective Hamiltonian cannot be neglected \emph{a priori}.

In the following, we will analyze numerical and experimental data, looking for evidence for the presence of $k'_a$, and comparing the relative weight of the different  contributions to $K'_a$.

\subsubsection*{On the value of $K''_a$}
We have obtained $K''_a=\kappa_0/4$ (see Eq.~\ref{Ksa}), where $\kappa_0$ is the bending rigidity of a symmetric membrane such that $\delta=0$. The elastic constant $\kappa_0$ is related to the bending rigidity $\kappa$ of the Helfrich model (see Methods, Sec.~\ref{special_cases}) through
\begin{equation}
\kappa=\kappa_0 -\frac{\kappa_0^2}{K_a}(c_0-c'_0\Sigma_0)^2\,.
\end{equation}
The difference between $\kappa_0$ and $\kappa$ arises from integrating out $\delta$ (see Methods, Sec.~\ref{elim_deltat}). In the previous models, this procedure was not carried out, as one focused directly on the symmetric case $\delta=0$. All previous models thus made the approximation $K''_a=\kappa/4$~\cite{Huang86, Goulian98, Nielsen98, Dan93, ArandaEspinoza96, Brannigan06}.

In addition, in Sec.~\ref{elim_deltat} of our Methods part, we show that $\zeta$ is also involved in $K''_a$, which stresses further the possible importance of such terms in order to describe membrane elasticity at the nanoscale.

\subsubsection*{On boundary terms}
The boundary terms correspond to the last three terms in Eq.~\ref{jbarc2b}. When one wishes to describe the local membrane deformation due to a transmembrane protein, boundary terms play an important part, as their integral on the contour of the protein contributes to the deformation energy. The first two boundary terms are the same as in Refs.~\cite{Dan93, ArandaEspinoza96, Brannigan06}. However, even at vanishing applied tension, we have $K'_a\neq-2\,A_2$, contrary to the previous models~\cite{Brannigan06}, due to the presence of $k'_a$. We have also accounted for the Gaussian bending rigidity $\bar\kappa$, as in Ref.~\cite{Brannigan07}: it yields the third boundary term. 

Again, the situation is more complex when $\beta$ is included, as the expressions of $A_1$ and $A_2$ then feature extra terms linear in $\beta$ (see Eq.~\ref{A1A2beta} in Sec.~\ref{elim_deltat} of our Methods part).

\subsubsection*{On lipid tilt}

Several membrane models including lipid tilt in addition to average shape deformations and/or thickness deformations have been elaborated~\cite{Fournier99, May99, May07, May07b, Watson11, Watson12}. These models provide improvements with respect to the Helfrich model, yielding better agreement with numerical data on bulk membranes~\cite{Watson11, Watson12}. 

Our model does not include lipid tilt because we focus on local thickness deformations, and especially on comparison to experimental and numerical data regarding deformations induced by mismatched proteins. While it would be interesting to include this extra degree of freedom, it would imply introducing several membrane parameters, which would make comparison to mismatch data impractical. 

Not taking tilt into account means that we are effectively integrating out this degree of freedom through coarse-graining. More precisely, the elastic coefficients of a more detailed membrane model, which would include tilt as an extra degree of freedom, would be renormalized by integrating out tilt. This means that tilt is included within the elastic coefficients of our membrane model. In addition, the interaction energy between the membrane and a mismatched inclusion (see, e.g., Eq.~\ref{potint}), and, consequently, the effective boundary conditions at the inclusion boundary, may involve tilt (see, e.g., Ref.~\cite{Fournier99}). In this interaction energy, tilt can be integrated out in the same way as in the bulk membrane energy. Hence, we are not losing any part of the elastic energy by disregarding the tilt degree of freedom. However, it is not impossible that a model including tilt truncated at second order could prove more efficient (e.g., have a wider domain of validity at short wavelengths) than one truncated at the same order and disregarding tilt.

\subsection*{Comparison with numerical results}

As numerical simulations become more and more realistic, they start providing insight into the behavior of systems on the microscopic scale where direct experimental observation is difficult. Lipid membranes (with or without inclusions) are no exception. Over the last decade, several groups have simulated bilayer systems over length- and time-scales long enough to give access to the material constants relevant for nanoscale deformations. These simulations provide interesting tests for theoretical models describing membrane elasticity at the nanoscale. We will compare the predictions of our model to recent numerical results in this Section. All the numerical results we will discuss have been obtained at zero applied tension. Hence, throughout this section, we take $\sigma=0$. This implies that our definition of the membrane thickness is equivalent to that considered in the original numerical works (see the discussion above on the definition of $u$).

\subsubsection*{Fluctuation spectra}
Using numerical simulations, one can measure precisely the fluctuation spectra of the average height and the thickness of a bilayer membrane~\cite{Lindahl00, Marrink01, Brannigan06, West09}. Microscopic protrusion modes, occurring at the scale of a lipid molecule, contribute to these spectra. While they are not described by continuum theories, it is possible to consider that they are decoupled from the larger-scale modes~\cite{Brannigan06, West09}. By fitting the numerical spectra to theoretical formulas, it is possible to extract the numerical values of the membrane constants involved in the continuum theory. In our framework, the fluctuation spectra of the average height of the membrane give access to the Helfrich bending rigidity $\kappa$, while those regarding the thickness of the membrane give access to $K_a$, $K'_a$ and $K''_a$. 

We have reanalyzed the height and thickness spectra presented in Refs.~\cite{Lindahl00, Marrink01, West09} using the fitting formulas in Refs.~\cite{Brannigan06, West09} (see Eq. 32 of Ref.~\cite{Brannigan06}) and the method described in Ref.~\cite{Brannigan06}, except that we did not assume that $K''_a=\kappa/4$, in order to include the possible effect of the difference between $\kappa$ and $\kappa_0$ (see Eq.~\ref{kappakappa0}), and of $\zeta$ (see Eq.~\ref{K''a_gen}). Our results were similar to those obtained in Refs.~\cite{Brannigan06, West09} assuming that $K''_a=\kappa/4$, and we obtained no systematic significant difference between $K''_a$ and $\kappa/4$, which means that the corrections to $K''_a$ predicted by our model are negligible in these simulations. This gives a justification for focusing only on the correction to $K'_a$, as we do in this article.
Besides, we obtained $K'_a<0$ from all the fits, as reported in Refs.~\cite{Brannigan06, West09}, and we checked that all the values obtained for $K'_a$ comply with the stability condition Eq.~\ref{stab}.

\subsubsection*{Deformation profiles close to a mismatched protein}

In Refs.~\cite{Brannigan06, Brannigan07, West09}, the thickness profile of a membrane containing one cylindrical inclusion with a hydrophobic mismatch has been obtained from coarse-grained numerical simulations. Comparing the average numerical thickness profiles to the equilibrium profiles predicted from theory is a good test for our model, in particular to find clues for the presence of $k'_a$. 

Let us denote the radius of the protein by $r_0$, and its hydrophobic length by $\ell$: the mismatch originates from the difference between $\ell$ and the equilibrium hydrophobic thickness $d_0$ of the membrane. The equilibrium shape of the membrane, which minimizes its deformation energy, is solution to the Euler-Lagrange equation associated with the effective Hamiltonian density in Eq.~\ref{jbarc2b}. We write down this equilibrium shape explicitly in Sec.~\ref{ThickProf} of our Methods part. In order to  determine it fully, it is necessary to impose boundary conditions at the edge of the inclusion, i.e., in $r=r_0$. There is a consensus on the assumption of strong hydrophobic coupling $u(r_0)=u_0=\ell-d_0$, as it costs more energy to expose part of the hydrophobic chains to water than to deform the membrane, for typical mismatches of a few \AA. Note that, with our definition of $u$, the condition $u(r_0)=u_0=\ell-d_0$ is valid to first order, while it is exactly valid with the definition of Refs.~\cite{Huang86, Goulian98, Nielsen98, Dan93, ArandaEspinoza96, Brannigan06, Brannigan07, Watson11} (see Eqs.~\ref{uprime},~\ref{ladiff}). This difference arises from the fact that our $u$ is not projected along $z$ (see Fig.~\ref{Dessin_defs}), which makes it fully independent of $h$. Given that the elastic energy is known to second order, the equilibrium membrane shape resulting from its minimization is known to first order, so it is sufficient to use boundary conditions to first order. Hence, such differences are not relevant for the present study and will not be mentioned any longer.

However, there is some debate about the second boundary condition in $r_0$ (see, e.g., Ref.~\cite{Brannigan06}), which regards the slope of the membrane thickness profile. 
Traditionally, one either assumes that the protein locally imposes a fixed slope to the membrane~\cite{Nielsen98, Goulian98}, or minimizes the effective Hamiltonian in the absence of any additional constraint, which amounts to considering that the system is free to adjust its slope in $r=r_0$~\cite{Dan93, ArandaEspinoza96, Brannigan06, Brannigan07, West09}. In Sec.~\ref{ThickProf} of our Methods part, we present the equilibrium profiles for these two types of boundary conditions. The actual boundary condition depends on the interactions between the protein and the membrane. In a quadratic approximation, these interactions generically give rise to an effective potential $f_s$ favoring a slope $s_0$ in $r_0$:
\begin{equation}
f_s=k_s\left(u'(r_0)-s_0\right)^2\,, \label{potint}
\end{equation}
where $k_s$ is an effective rigidity, while $u'$ denotes the derivative of the membrane thickness profile $u$ with respect to the radial coordinate $r$. Two \emph{a priori} unknown parameters, $k_s$ and $s_0$, are associated with this effective potential. The ``free-slope'' boundary condition (also called ``natural'' boundary condition~\cite{Dan93, Brannigan06}) is recovered in the limit $k_s\rightarrow 0$, which is appropriate if $f_s$ is negligible with respect to the energetic contributions in $f$. Conversely, if $k_s\rightarrow\infty$, the protein locally imposes the fixed slope $s_0$. If the interactions between the protein and the membrane lipids are sufficiently short-ranged, the protein cannot effectively impose or favor a slope on the coarse-grained membrane thickness profile. For instance, in the numerical simulations of Refs.~\cite{Brannigan06, Brannigan07, West09}, the interactions between the protein and the membrane lipids are of similar nature and of similar range as those between membrane lipids. Thus, we will choose the free-slope boundary condition in our analysis of this data. This choice was already made in Refs.~\cite{Brannigan06, Brannigan07, West09}. A practical advantage of this boundary condition is that it does not introduce any unknown parameter in the description.

\begin{figure}[ht]
\centerline{\includegraphics[width=0.5\textwidth]{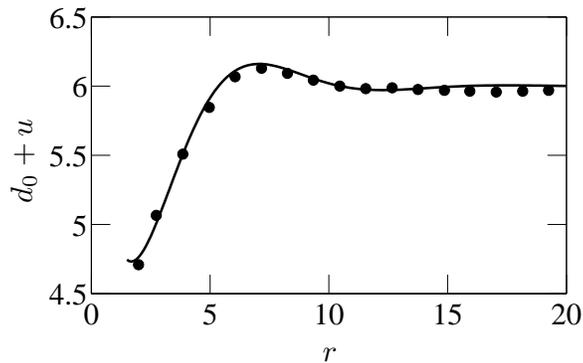}}
\caption{\textbf{Thickness deformation due to a mismatched inclusion.} Membrane thickness profile from Ref.~\cite{West09} in the vicinity of a mismatched inclusion with hydrophobic thickness $\ell\simeq2.4\,\mathrm{nm}$ and radius $r_0=9\,\textrm{\AA}$, with center in $r=0$, as a function of the radial coordinate $r$. The equilibrium membrane hydrophobic thickness is $d_0\simeq3.6\,\mathrm{nm}$. The unit of length on the graph is 6~\AA, as in Ref.~\cite{West09}. Dots: numerical data (the error bars on the data, not reproduced here, are about 1~\AA~wide~\cite{West09}). Line: best fit. Exactly as in the original reference, the numerical data is fitted to Eqs.~\ref{sol_fix}-\ref{Apm_b_c0} with $k'_a=0$, taking $u_0$ and the (renormalized) spontaneous curvature $\tilde c_0$ as fitting parameters, the other constants being known from the fluctuation spectra. \label{leprofil}}
\end{figure}

\begin{figure}[ht]
\centerline{\includegraphics[width=0.42\textwidth]{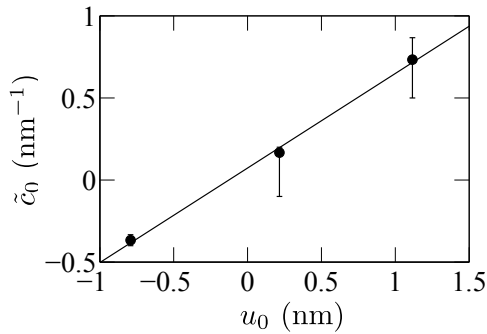}}
\caption{\textbf{Renormalized spontaneous curvature $\tilde c_0$ as a function of the hydrophobic mismatch $u_0$.} Data from Ref.~\cite{West09}, which presents fits of simulation results for inclusions with three different hydrophobic thicknesses. Line: linear fit, with slope $(0.56\pm0.02)\,\mathrm{nm}^{-2}$. Note that our $\tilde c_0$ corresponds to twice that in Table~2 of Ref.~\cite{West09}, as we work with total curvatures instead of average curvatures. The error bars on $\tilde c_0$ are those listed in that table, and $u_0$ corresponds to $2\,t_R^\mathrm{el}$ in that table. \label{West_lin}}
\end{figure}

The membrane model of Refs.~\cite{Brannigan06, Brannigan07, West09} is very similar to ours, except that $k'_a=0$. It was shown in Ref.~\cite{West09} that this model can reproduce very well the numerical results, \emph{provided that the spontaneous curvature is adjusted for each deformation profile} (see Fig.~\ref{leprofil}). In Ref.~\cite{West09}, the adjusted ``renormalized spontaneous curvature'', denoted by $\tilde c_0$, was found to depend linearly on the hydrophobic mismatch $u_0$ \cite{West09}, as shown in Fig.~\ref{West_lin}. In our model, the equilibrium profile corresponding to the free-slope boundary conditions (see Eqs.~\ref{sol_fix} and~\ref{Apm_b_c0}) involves $k'_a$. We show in Sec.~\ref{ThickProf} of our Methods part that the quantity 
\begin{equation}
\tilde c_0=c_0+\frac{k'_a}{\kappa}u_0\,,
\label{c0u0}
\end{equation}
then plays the part of the renormalized spontaneous curvature of Ref.~\cite{West09} in the equilibrium profile. This quantity is linear in $u_0$: our model, and more precisely the presence of a nonvanishing $k'_a$, thus provides an appealing explanation for the linear dependence observed in Ref.~\cite{West09}. 

Using a linear fit of the data of Ref.~\cite{West09} (see Fig.~\ref{West_lin}), together with Eq.~\ref{c0u0} and the value $\kappa=2.3\times 10^{-20}\,\mathrm{J}$ extracted from the spectra in Ref.~\cite{West09}, we obtain $k'_a=13\,\mathrm{mN/m}$. 

It is interesting to compare this value to $K'_a$, which is obtained from the fluctuation spectra in Ref.~\cite{West09}: $K'_a=-9.2\,\mathrm{mN/m}$. This shows that the contribution of $k'_a$ to $K'_a$ is important. Besides, we may now estimate the contribution to $K'_a$ that stems from the monolayer spontaneous curvature (see Eq.~\ref{Kpa}): $-\kappa_0(c_0-c'_0\Sigma_0)/d_0=K'_a-k'_a=-22\,\mathrm{mN/m}$. Using values from the fluctuation spectra in Ref.~\cite{West09}, this yields $\xi\approx -6\,\mathrm{\AA}$ for the algebraic distance from the neutral surface of a monolayer to the hydrophilic-hydrophobic interface of this monolayer (see Methods, Sec.~\ref{appcp0} for the relation between $\xi$ and $c'_0$).

\begin{figure}[htb]
\centerline{\includegraphics[width=0.53\textwidth]{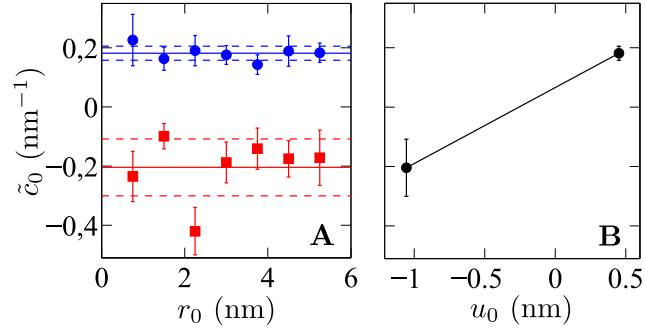}}
\caption{\textbf{Renormalized spontaneous curvature $\tilde c_0$ as a function of the inclusion radius $r_0$ and the hydrophobic mismatch $u_0$.} \textbf{A)} $\tilde c_0$ versus $r_0$. The values of $\tilde c_0$ were obtained by fitting each thickness deformation profile of Ref.~\cite{Brannigan07}. Circles (blue): positive mismatch, $u_0=0.45\, \mathrm{nm}$. Squares (red): negative mismatch, $u_0=-1.1\, \mathrm{nm}$. Solid lines: average values; dotted lines: standard deviation over the seven data points (corresponding to the different $r_0$), for each value of $u_0$. \textbf{B)} Average value of $\tilde c_0$ (see A) as a function of the hydrophobic mismatch $u_0$. The equation of the line joining the two data points has a slope $0.26\,\mathrm{nm}^{-2}$.\label{Brannigan_r_u0}}
\end{figure}

In Ref.~\cite{Brannigan07}, a different coarse-grained molecular simulation model was used to obtain the equilibrium membrane thickness profiles for cylindrical inclusions with two different hydrophobic thicknesses, one yielding a positive mismatch and the other a negative one, and with seven different radii $r_0$. These profiles are presented in Figs.~6 and~7 of Ref.~\cite{Brannigan07}, except those corresponding to the inclusions with largest radii (5.25 nm), but this data was communicated to us by one of the authors of Ref.~\cite{Brannigan07}. We fitted each of these numerical profiles to the analytical equilibrium profile Eq.~\ref{sol_fix} with prefactors $A_\pm(0,\tilde c_0)$ (see Eq.~\ref{corresp}), using $\tilde c_0$ as our only fitting parameter, in the spirit of Ref.~\cite{West09}. We found that $\tilde c_0$ does not depend on the radius of the inclusion, but that it depends significantly on the mismatch (see Fig.~\ref{Brannigan_r_u0}A). This is in good agreement with the predictions of our model (see Eq.~\ref{c0u0}). For each of the two values of $u_0$, we have averaged $\tilde c_0$ over the seven results corresponding to the different inclusion radii. The line joining these two average values of $\tilde c_0$ as a function of $u_0$ is plotted in Fig.~\ref{Brannigan_r_u0}B. Using Eq.~\ref{c0u0} and the value $\kappa=1.4\times 10^{-19}\,\mathrm{J}$~\cite{Brannigan06, Brannigan07}, the slope of this line yields $k'_a=36\,\mathrm{mN/m}$: the order of magnitude of this value is the same as the one obtained from the data of Ref.~\cite{West09}.

Again, we can compare this value to $K'_a$, which is obtained from the fluctuation spectra in Refs.~\cite{Brannigan06, Brannigan07}: $K'_a=-11.9\,\mathrm{mN/m}$. Hence, the contribution of $k'_a$ to $K'_a$ is important here too. We also obtain $-\kappa_0(c_0-c'_0\Sigma_0)/d_0=K'_a-k'_a=-48\,\mathrm{mN/m}$, and $\xi\approx -3\,\mathrm{\AA}$. 

\begin{figure}[htb]
\centerline{\includegraphics[width=0.45\textwidth]{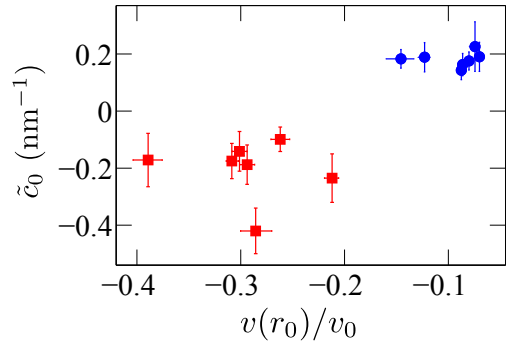}}
\caption{\textbf{Renormalized spontaneous curvature $\tilde c_0$ versus the relative volume variation $v(r_0)/v_0$ on the inclusion edge.} The values of $\tilde c_0$ are extracted from fitting the data of Ref.~\cite{Brannigan07}, and the values of $v(r_0)/v_0$ are directly taken from Ref.~\cite{Brannigan07}. \label{Brannigan_v}}
\end{figure}

In Ref.~\cite{Brannigan07}, the shortcomings of the model that disregards $k'_a$ are explained by the local variation of the volume per lipid close to the protein. It is shown in Ref.~\cite{Brannigan07} that including this effect yields 
\begin{equation}
\tilde c_0=c_0-\frac{\eta}{v_0}v(r_0)\,,
\end{equation}
where $v_0$ is the bulk equilibrium volume per lipid, while $v_0+v(r_0)$ denotes the volume per lipid in $r=r_0$. However, the predicted linear dependence of $\tilde c_0$ in $v(r_0)/v_0$ is not obvious: in Fig.~\ref{Brannigan_v}, we rather see two groups of points (one for each value of $u_0$) than a linear law. In other words, the data of Ref.~\cite{Brannigan07} is more consistent with a value of $\tilde c_0$ that depends only on $u_0$ and not on $v$ (or $r_0$), in agreement with the predictions of our model (see Eq.~\ref{c0u0}). In Ref.~\cite{West09}, local modifications of the volume per lipid close to the inclusion were investigated too, as well as local modifications of the nematic order, of the shielding of the hydrophobic membrane interior from the solvent, and of the overlap between the two monolayers. None of these effects was found to explain satisfactorily the linear dependence of $\tilde c_0$ versus $u_0$~\cite{West09}.

To sum up, our model can explain the dependence of $\tilde c_0$ in $u_0$ observed in the numerical results of Refs.~\cite{Brannigan07, West09} as a consequence of the presence of $k'_a$. Our explanation does not involve any local modification of the membrane properties, in contrast with those proposed in Refs.~\cite{Brannigan07, West09}. Furthermore, the order of magnitude we obtain for $k'_a$ from the data of Refs.~\cite{West09, Brannigan07} is in agreement with our estimate in Eq.~\ref{kpa_g}.

\subsection*{Comparison with experimental results}

The antimicrobial linear pentadecapeptide gramicidin (see~\cite{Kelkar07} for a review) is a very convenient experimental system to probe membrane elasticity on the nanoscale. In lipid membranes, two gramicidin monomers (one in each monolayer) associate via the N-terminus to form a dimeric channel, stabilized by six intermolecular hydrogen bonds. The channel being large enough for the passage of monovalent cations, conductivity measurements~\cite{OConnell90} can detect its formation and lifetime, which are directly influenced by the membrane properties. Indeed, while the monomers do not deform the membrane, the dimeric channel presents a hydrophobic mismatch with the membrane, so that dimer formation involves a local deformation of the bilayer. The gramicidin channel can therefore act as a local probe for the bilayer elasticity. Furthermore, the gramicidin channel can be considered as up-down symmetric and cylinder-shaped, which makes it convenient for theoretical investigations.

Data on gramicidin channels originally motivated theoretical investigations on membrane models describing local thickness deformations~\cite{Huang86,Helfrich90,Dan93,ArandaEspinoza96}. Such data now provides a great opportunity to test any refinement of these models. We will compare our model to the data of Ref.~\cite{Elliott83} regarding the lifetime of the gramicidin channel as a function of bilayer thickness, and then to the data of Ref.~\cite{Goulian98} regarding the formation rate of the gramicidin channel as a function of bilayer tension.

In order to compare the predictions of our model to experimental data regarding the gramicidin channel, it is necessary to make some assumptions about the boundary conditions at the edge of the channel, i.e., in $r=r_0$. As discussed in the previous section, we will assume strong hydrophobic coupling, i.e., $u(r_0)=u_0=\ell-d_0$, but determining the boundary condition on the slope of the membrane thickness profile is trickier as it depends on the interactions between gramicidin and the membrane lipids. In previous analyses~\cite{Goulian98, Lundbaek99}, the fixed-slope boundary condition was favored as giving the best agreement with experimental data. However, different values of the fixed slope were obtained in these studies. In addition, recent all-atom simulations of gramicidin channels in lipid bilayers indicate that the membrane thickness profile is complex in the first lipid shell around the channel, due to specific interactions, and that beyond this first shell, no common slope exists for the different membranes investigated~\cite{Kim12}. Given the difficulty to determine the actual effective boundary condition associated with the slope of the membrane thickness profile, we will adopt the free-slope boundary condition, which has the advantage not to introduce any unknown parameter in the analysis, but we will also compare our results to those obtained with the more traditional fixed-slope boundary condition.

\subsubsection*{Analysis of the experimental data of Elliott \textit{et al.} \cite{Elliott83}}\label{subsec:Elliott}

It was shown in Ref.~\cite{Nielsen98} that the detailed elastic membrane model introduced in Ref.~\cite{Huang86} yields an effective linear spring model as far as the membrane deformation due to gramicidin is concerned~\cite{Nielsen98,Lundbaek99}: the energy variation $F$ associated with the deformation can be expressed as $F=Hu_0^2$, where $H$ is the effective spring constant, while $u_0$ is the thickness mismatch between the gramicidin channel and the membrane. This linear spring model was validated by comparison with experimental data on the lifetime of the gramicidin channel, measured as a function of bilayer thickness (\cite{Kolb77,Elliott83}, summarized in \cite{Lundbaek99}) and as a function of the channel length \cite{Hwang03}.

We will here focus on the data concerning virtually solvent-free bilayers, i.e., membranes formed using squalene. The elasticity of membranes containing hydrocarbons should be different: for instance, a local thickness change of the membrane could be associated with a redistribution of the hydrocarbons. (In this, our analysis differs from that of Ref.~\cite{Brannigan06}, where all the data of Ref.~\cite{Elliott83} was considered. Another important difference with the analysis conducted in that reference is that we use experimental values of the membrane parameters, which are quite different from the values coming from numerical simulations.) In Ref.~\cite{Lundbaek99}, the effective spring constant $H$ of the membrane was estimated from data of Ref.~\cite{Elliott83} on gramicidin channel lifetime for three bilayers formed in squalene with monoglycerids that differed only through their chain lengths: the different thicknesses of these membranes yield different hydrophobic mismatches with a given type of gramicidin channels. The value $H_\mathrm{exp}=115\pm 10\,\,\mathrm{mN.m^{-1}}$ was obtained.

In Sec.~\ref{defen} of our Methods part, we use our model to calculate the deformation energy of the membrane due to the presence of a mismatched protein. Both in the case of the free-slope boundary condition, and in the case where the gramicidin channel locally imposes a vanishing slope, this deformation energy can be expressed as a quadratic function of the mismatch $u_0$. The prefactor of $u_0^2$ in the deformation energy $F$ corresponds to the effective spring constant of the system. Thus, although our model is different from the one of Refs.~\cite{Huang86,Nielsen98,Goulian98}, it also yields an effective linear spring model. This is not surprising since we are dealing with the small deformations of an elastic system. However, the detailed expressions of our spring constants as a function of the membrane parameters (see Eqs.~\ref{ressortfixe} and~\ref{ressortlibre}) are different from those obtained using the model of Refs.~\cite{Huang86,Nielsen98,Goulian98}, due to the differences between the underlying membrane models. In particular, in our model, $k'_a$ is involved in $H$, through $K'_a$. Our aim will be to find out which value of $k'_a$ gives the best agreement with the experimental value of $H$.

Using Eqs.~\ref{Kpa},~\ref{Ksa} and~\ref{A2}, and neglecting the difference between $\kappa$ and $\kappa_0$, Eqs.~\ref{ressortfixe} and~\ref{ressortlibre} show that $H$ depends on the elastic constants $\kappa$, $\bar\kappa$ and $c_0$ involved in the Helfrich model, on $K_a$, on $c'_0\Sigma_0$, which corresponds to the spontaneous curvature variation with the area per lipid, on $d_0$, on the radius $r_0$ of the gramicidin channel, and on $k'_a$. There is, to our knowledge, no direct experimental measurement of $c'_0\Sigma_0$ available, but, as shown in Sec.~\ref{appcp0} our Methods part, we have $c'_0\Sigma_0=K_a\xi/\kappa$, where $\xi$ denotes the algebraic distance from the neutral surface of a monolayer to the hydrophilic-hydrophobic interface of this monolayer (see Eq.~\ref{c0primeexpr}, neglecting the difference between $\kappa$ and $\kappa_0$). Hence, in order to calculate the spring constant, we need values for $\kappa$, $\bar\kappa$, $c_0$, $K_a$ and $\xi$, in the precise case of monoolein membranes.

In Ref.~\cite{Chung94Nat}, the elastic constants $\kappa$, $\bar\kappa$ and $c_0$ were measured in a monoolein cubic mesophase, both at $25^\circ\mathrm{C}$ and at $35^\circ\mathrm{C}$. The positions of the neutral surface and of the hydrophilic-hydrophobic interface were estimated on the same system in Ref.~\cite{Chung94BJ}, but these results were flawed by a mathematical issue, which was corrected in Ref.~\cite{Templer95L}. This correction yielded other corrections on $c_0$, and on the ratio $\bar \kappa/\kappa$~\cite{Templer95JPII}. These results regard a cubic phase, where the membrane is highly deformed with respect to a flat bilayer: the values of the various constants should be affected by the strains present in this phase. In another work~\cite{Vacklin00}, the constants of monoolein are determined in a highly hydrated doped $\mathrm{H}_{\mathrm{II}}$ phase, where the strains are better relaxed. However, these measurements were carried out at $37^\circ\mathrm{C}$, while the experiments of Ref.~\cite{Elliott83} that we wish to analyze were performed at $23^\circ\mathrm{C}$. Given that the data of Refs.~\cite{Chung94Nat,Chung94BJ} include the most appropriate temperature, while the ones of Ref.~\cite{Vacklin00} correspond to the most appropriate phase, we will present results corresponding to both sets of parameters. Finally, the experimental value of $K_a$ for monoolein is provided by Ref.~\cite{Hladky82}. 

In Table~\ref{tableH}, we present the results obtained for the spring constant $H$ of monoolein bilayers, using the different experimental estimates of the membrane constants. The main difference between parameter sets 1 and 2 is the value and the sign of $\bar\kappa$~\cite{Chung94Nat,Templer95JPII}. However, $\bar\kappa$ is involved in $H$ only in the free-slope case (see Eqs.~\ref{ressortfixe} and~\ref{ressortlibre}): the 3\% difference between the values of $H_0$ obtained with parameter sets 1 and 2 stems only from the difference on $c_0$, while the 12\% difference between $H_f$ obtained with data sets 1 and 2 contains an important contribution from $\bar\kappa$. The constants in parameter set 3, corresponding to Ref.~\cite{Vacklin00}, are significantly different from those of Refs.~\cite{Chung94Nat,Templer95JPII}, which yields a 30\% difference on $H_0$ and a 20\% difference on $H_f$. We also note that, as the value of the algebraic distance from the neutral surface to the hydrophilic-hydrophobic interface of a monolayer is very small compared to the other length scales involved ($\xi=-0.3\,\mathrm{\AA}$~\cite{Templer95L}), the contribution of $c'_0\Sigma_0$ to $H$ is negligible (it is of order 1\%). 

Let us now discuss the results given by our model, in the case of the free-slope boundary condition (see Table~\ref{tableH}). The spring constants $H_f$ obtained assuming that $k'_a=0$ are about three times smaller than the experimental value $H_\mathrm{exp}=115\pm 10\,\,\mathrm{mN.m^{-1}}$ (see line 1 of Table~\ref{tableH}). (This result is very similar to that in Ref.~\cite{Lundbaek99}, which illustrates that accounting for monolayer spontaneous curvature and for boundary terms does not change much the value of $H$.) However, $H_f$ reaches the experimental value for $k'_a\approx25\,\mathrm{mN/m}$ for all three parameter sets (see line 2 of Table~\ref{tableH}). Hence, for free-slope boundary conditions, the presence of $k'_a$, with an order of magnitude consistent with Eq.~\ref{kpa_g}, improves the agreement between theory and experiment. 

We may compare these values of $k'_a$ to the contribution to $K'_a$ that originates from the monolayer spontaneous curvature (see Eq.~\ref{Kpa}): $-\kappa_0(c_0-c'_0\Sigma_0)/d_0$. We estimate the value of this contribution to be between $0.26$ and $1.2\,\mathrm{mN/m}$, depending on which set of parameters is chosen. This is positive and much smaller in absolute value than the estimates obtained from the numerical data of Ref.~\cite{West09} and of Ref.~\cite{Brannigan07}: here, the neutral surface of a monolayer and its hydrophilic-hydrophobic interface are very close, while $\xi$ seemed to be significant in the numerical simulations. In addition, the contribution of membrane tension to $K'_a$, namely, $\sigma/4$, cannot exceed about 1 mN/m. In the case of the free-slope boundary condition, our results imply that $k'_a$ should be the dominant contribution to $K'_a$ for the membranes studied in Ref.~\cite{Elliott83}.

Let us now discuss the results obtained for the zero-slope boundary condition, which was investigated in Ref.~\cite{Lundbaek99}. For the zero-slope boundary condition, the values obtained for $H_0$ assuming that $k'_a=0$ are in quite good agreement with the experimental value $H_\mathrm{exp}=115\pm 10\,\,\mathrm{mN.m^{-1}}$ obtained in Ref.~\cite{Lundbaek99} from the data of Ref.~\cite{Elliott83}, for all the data sets we used (see line 3 of Table~\ref{tableH}): hence, $k'_a$ seems negligible if zero-slope boundary conditions are assumed. However, there is no justification to assume that the gramicidin channel locally imposes a vanishing slope.

\subsubsection*{Analysis of the experimental data of Goulian \textit{et al.} \cite{Goulian98}}\label{subsec:Goulian}

While the experiments cited in the previous Section dealt with discrete changes of the hydrophobic mismatch obtained by varying membrane composition, Goulian \textit{et al.}~\cite{Goulian98} measured the gramicidin channel formation rate $f$ in lipid vesicles as a function of the tension $\sigma$ applied through a micropipette. As the tension is an externally controlled parameter that can be changed continuously for the same gramicidin-containing membrane, this approach can yield more information, and it has the advantage of limiting the experimental artifacts associated to new preparations. To date, the experiment in Ref.~\cite{Goulian98} remains the most significant in the field and should serve as a testing ground for any theoretical model. We will therefore discuss in detail the data and its interpretation by the original authors~\cite{Goulian98,Nielsen98} as well as in terms of our model (see Eq.~\ref{jbarc2b}).

Within experimental precision, the data of Ref.~\cite{Goulian98} can be described by a quadratic dependence:
\begin{equation}
\ln f  = g(\sigma) = C_0 + C_1 \sigma + C_2 \sigma ^2.
\label{eq:quadratic}
\end{equation}
Given that $\ln f$ is a linear function of the energy barrier associated with the formation of the gramicidin dimer, it is a sum of a chemical contribution, including, e.g., the energy involved in hydrogen bond formation, and of a contribution arising from membrane deformation due to the dimer (monomers do not deform the membrane)~\cite{Goulian98}. The latter contribution arises from the hydrophobic mismatch between the membrane and the dimer, and it depends on the applied tension $\sigma$, since the membrane hydrophobic thickness depends on $\sigma$ (see Eq.~\ref{deq} in Sec.~\ref{Tension} of our Methods part). Expressing the deformation energy $F$ of the membrane due to the presence of the dimer gives a theoretical expression for the $\sigma$-dependent part of $\ln f$. In our model, $K'_a$ features a contribution coming from $\sigma$ (see Eq.~\ref{Kpa}). However, this term is negligible, given that $\sigma/4\ll \sqrt{K_a K''_a}/d_0$ (see Eq.~\ref{negl}), for realistic tension values (a few mN/m at most), and for the experimentally measured values of the membrane constants~\cite{Rawicz00}. This enables us to disregard it. Then, our quadratic elastic membrane model simply gives a quadratic dependence of $F$ on $\sigma$, in agreement with the form of Eq.~\ref{eq:quadratic}. Comparing the experimental values of $C_1$ and $C_2$ to those predicted by theory provides a test for theoretical models~\cite{Goulian98}. (Note that, if the $\sigma$-dependent contribution to $K'_a$ is included, the expression of the $\sigma$-dependent part of $\ln f$ is no longer simply quadratic in $\sigma$. However, we explicitly verified that including this contribution yields a negligible change to the relation between $\sigma$ and $\ln f$, for realistic values of the parameters). 

\begin{figure}[htb]
\centerline{\includegraphics[width=0.6\textwidth]{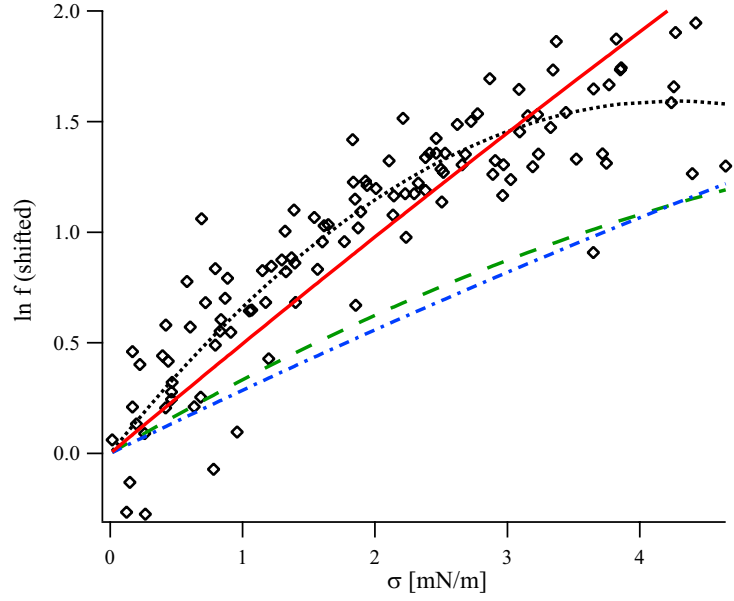}}
\caption{\textbf{Formation rate $f$ of gramicidin channels versus the applied tension $\sigma$, analyzed with a quadratic model.} Diamonds: experimental data retrieved from Fig.~6b of Ref.~\cite{Goulian98}, after subtraction of the baselines $C_0^k$. Dotted black line: best quadratic fit, with $C_1 = 0.74 \,\mathrm{(mN/m)}^{-1}$ and $C_2 = -0.09 \,\mathrm{(mN/m)}^{-2}$; $\chi^2\equiv\chi^2_\mathrm{min}$. Dashed green line: results obtained from the elastic model of Ref.~\cite{Nielsen98}, with the constants given in~\cite{Goulian98}; $\chi^2/\chi^2_\mathrm{min}=5.72$. Dashed-dotted blue line: \emph{idem} with more recent values of the constants; $\chi^2/\chi^2_\mathrm{min}=6.68$. Solid red line: results obtained by taking $s=0$ and the recent values of the constants in the model of Refs.~\cite{Huang86, Nielsen98}; $\chi^2/\chi^2_\mathrm{min}=1.75$. The values of $C_1$ and $C_2$ corresponding to the curves on this graph are listed in Table~\ref{tableC1C2_1}. }
\label{fig:quadratic_fit}
\end{figure}

Since the coefficients $C_1$ and $C_2$ arise from membrane elasticity, they are common to all the vesicles studied in Ref.~\cite{Goulian98}, which have the same lipid composition. Conversely, the baseline $C_0$ depends on parameters such as the concentration of gramicidin molecules, so it can take a different value for each of the twelve vesicles studied in Ref.~\cite{Goulian98}. A global fit to the data of Ref.~\cite{Goulian98} using Eq.~\ref{eq:quadratic} involves minimizing the goodness-of-fit function 
\begin{equation}
\chi ^2 = \sum _j (\ln f_j - g(\sigma _j))^2\,,
\label{chisq_form}
\end{equation}
where the index $j$ runs over all the experimental points, with fitting parameters $C_1, C_2, C_0^k, \, k=1, \dots, 12$. The baseline $C_0^k$ is then subtracted from each of the twelve curves. All the data is plotted in the same graph in Fig.~\ref{fig:quadratic_fit}. The best global fit, corresponding to $C_1 = 0.74 \pm 0.07 \,\mathrm{(mN/m)}^{-1}$ and $C_2 = -0.090 \pm 0.015 \,\mathrm{(mN/m)}^{-2}$, is shown on Fig.~\ref{fig:quadratic_fit} as the dotted (black) line. (It should be noted that the values obtained by fitting the individual curves are much more scattered: $C_1$ ranges from $0.4$ to $1.5 \,\mathrm{(mN/m)}^{-1}$ and $C_2$ from $-0.3$ to $0 \,\mathrm{(mN/m)}^{-2}$.)

\begin{figure}[htb]
\centerline{\includegraphics[width=0.6\textwidth]{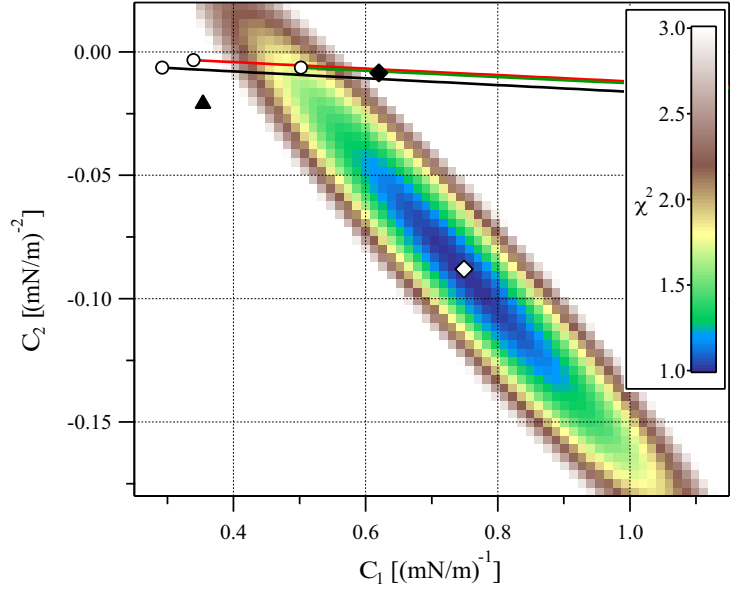}}
\caption{\textbf{Comparison between the experimental values of $C_1$ and $C_2$ and those obtained from different models.} Colorscale: goodness-of-fit function $\chi ^2$ (see Eq.~\ref{chisq_form}) for the data of Ref.~\cite{Goulian98}, as a function of the fitting parameters $C_1$ and $C_2$. White diamond: values of $C_1$ and $C_2$ that give the best fit. Black triangle: results obtained from the elastic model of Ref.~\cite{Nielsen98}, with the constants given in~\cite{Goulian98}. Lines: trajectories obtained from our model in the $(C_1, C_2)$ plane when varying $k'_a$. Red: free slope; green: $s=0$, black: $s=0.3$. These three curves start by a white dot at $k'_a=0$, and $k'_a$ increases rightwards along these curves. The rightmost white dot ($k'_a=0$, $s=0$) roughly corresponds to the best agreement we can obtain between our model and the experiment fitted to the quadratic model (red curve on Fig.~\ref{fig:quadratic_fit}). The black diamond corresponds to the best agreement we can obtain between our model and the experiment fitted to the linear model at low tension (see Fig.~\ref{fig:linear_fit}).}
\label{fig:Chisq}
\end{figure}

In Ref.~\cite{Goulian98}, the authors used published values of the material constants to calculate $C_1$ and $C_2$ in the framework of their elastic model~\cite{Nielsen98}, based on that of Ref.~\cite{Huang86}. Using fixed-slope boundary conditions, they reported good agreement with the experimental data for a reasonable value of the unknown slope $s$ ($s=0.3$). However, we need to raise the following points:
\begin{enumerate}
\item{} There was a mistake in their implementation of the formula of Ref.~\cite{Nielsen98} giving $C_1$ and $C_2$ as a function of the material constants. More precisely, we found that a factor of 2 was missing in the expression of $C_1$ and a factor of 4 was missing in that of $C_2$ in the implementation of the formula of Ref.~\cite{Nielsen98}. This was confirmed by Mark Goulian (private communication). The actual values of $C_1$ and $C_2$ obtained using the same values of the constants as in Ref.~\cite{Goulian98} are in fact quite far from those corresponding to the best fit of the experimental data, as shown by the dashed green line in Fig.~\ref{fig:quadratic_fit} (see also Fig.~\ref{fig:Chisq} and Table~\ref{tableC1C2_1}).
\item{} The estimates for the elastic constants used in Ref.~\cite{Goulian98} are somewhat different from more recent and more widely accepted values. Henceforth, we will use the following parameters, for a DOPC membrane: $d_0= 2.7\,\mathrm{nm}$~\cite{Goulian98}, $K_a=265\,\mathrm{mN/m}$, $\kappa=8.5\times10^{-20}\,\mathrm{J}$~\cite{Rawicz00}, $c_0=-0.132\,\mathrm{nm}^{-1}$~\cite{Szule02}, and the dimensions of a gramicidin channel: $r_0=1\,\mathrm{nm}$, $\ell'=\ell+\delta=2.3\,\mathrm{nm}$~\cite{Goulian98}. Implementing these more recent values in the model of Ref.~\cite{Nielsen98} does not yield a better agreement with experiment, as shown by the dashed-dotted (blue) line in Figure~\ref{fig:quadratic_fit} (see also Fig.~\ref{fig:Chisq} and Table~\ref{tableC1C2_1}). 
\end{enumerate}
A somewhat better agreement with the experimental data is obtained when taking $s=0$ instead of $s=0.3$ for the fixed slope (see Figs.~\ref{fig:quadratic_fit} and~\ref{fig:Chisq}, and Table~\ref{tableC1C2_1}). However, the downward inflection of the experimental curves at high $\sigma$ is not adequately described for any value of $s$. In fact, $C_2$ is independent of $s$, and its absolute value given by the elastic model is 15 times smaller than the experimental one (see Table~\ref{tableC1C2_1}). We conclude that the elastic model of Refs.~\cite{Huang86,Nielsen98} does not satisfactorily describe the data of Ref.~\cite{Goulian98} regarding the lifetime of the gramicidin channel under tension. 

In Sec.~\ref{defen} of our Methods part, we calculate the deformation energy $F$ in the framework of our model, both for the fixed-slope boundary condition and for the free-slope boundary condition. The resulting expressions of $C_1$ and $C_2$ are given by Eqs.~\ref{C1fixe},~\ref{C2fixe}, \ref{C1libre} and~\ref{C2libre}. In order to see which values of $k'_a$ and which boundary conditions give the best agreement with the experiments of Ref.~\cite{Goulian98}, we present a plot of the goodness-of-fit function $\chi ^2$ (see Eq.~\ref{chisq_form}) in a $(C_1,C_2)$ graph in Fig.~\ref{fig:Chisq}. On this graph, we have plotted the trajectories obtained from our model in the $(C_1, C_2)$ plane when varying $k'_a$, for $s=0$, for $s=0.3$ (as in Ref.~\cite{Goulian98}), and for the free-slope boundary condition.

In order to obtain numerical values of $C_1$ and $C_2$ from Eqs.~\ref{C1fixe},~\ref{C2fixe},~\ref{C1libre} and~\ref{C2libre}, we used the above-mentioned parameter values, and the estimate $\bar\kappa= -0.8\kappa$~\cite{West09}. Finally, we estimated $c'_0\Sigma_0$ through the relation $c'_0\Sigma_0=K_a\xi/\kappa$ (see Eq.~\ref{c0primeexpr} in Sec.~\ref{appcp0} of our Methods part). For this, the algebraic distance $\xi$ from the neutral surface of a monolayer to the hydrophilic-hydrophobic interface of this monolayer was estimated by first determining the position of the pivot surface from the data of Ref.~\cite{Szule02}, and by calculating the distance between it and the neutral surface~\cite{Leikin96}: we found $\xi\approx-0.5\,\mathrm{\AA}$. Here again, the neutral surface is close to the hydrophilic-hydrophobic interface. For the sake of simplicity, we took $c'_0=0$, and we checked that the results were not significantly different when taking $\xi=-0.5\,\mathrm{\AA}$.

The ingredient in our model that can change significantly the results is $k'_a$ (Note that the values of $C_1$ and $C_2$ corresponding to $k'_a=0$ are very close to those obtained using the model of Ref.~\cite{Goulian98} with our values of the parameters, as shown in Table~\ref{tableC1C2_1}. This illustrates again that the influence of boundary terms is quantitatively small.) Fig.~\ref{fig:Chisq} shows that the experimental value of $C_1$ can be explained by our model. In addition, the values of $k'_a$ that minimize $\chi^2$, i.e., that give the best agreement with the experimental data of Ref.~\cite{Goulian98}, are between $0$ and $50\,\mathrm{mN/m}$, depending on the boundary condition chosen, as shown in Table~\ref{tableC1C2_0}. This range of values of $k'_a$ is reasonable.  

For the free-slope boundary condition, the best agreement with the experimental results is obtained for $k'_a\approx40\,\mathrm{mN/m}$ (see Table~\ref{tableC1C2_0} and Fig.~\ref{fig:Chisq}). The order of magnitude is the one expected from $k'_a\approx\gamma/2$. 

Let us now discuss the results obtained for the fixed-slope boundary condition, which is used in Ref.~\cite{Goulian98}. For a fixed slope $s=0$, the best agreement with the results of Ref.~\cite{Goulian98} analyzed with the complete quadratic fit is obtained for $k'_a=0$. Conversely, for $s=0.3$, the best agreement is obtained for $k'_a\approx40\,\mathrm{mN/m}$, which is similar to the result obtained the free-slope case (see Table~\ref{tableC1C2_0} and Fig.~\ref{fig:Chisq}). Hence, in the case of the fixed-slope boundary condition, the conclusions depend a lot on the value of $s$ that is chosen. 

In all cases, the absolute values of $C_2$ we obtain remain much smaller than the one that matches best the experimental results, which is $C_2 = -90.0\times 10^{-3}\,\mathrm{(mN/m)}^{-2}$ (see Fig.~\ref{fig:quadratic_fit}). This can be seen in Fig.~\ref{fig:Chisq}, as well as in Table~\ref{tableC1C2_0}. Hence, with our model, as with the one of Ref.~\cite{Goulian98}, it seems impossible to explain the experimental value of $C_2$. Our model predicts that $C_2$ is proportional to the effective spring constant $H$ of the membrane discussed in the previous Section (see Eqs.~\ref{ressortfixe} and~\ref{ressortlibre}): it is thus quite unexpected to have a good agreement with the experimental values of $H$ but not with those of $C_2$. This disagreement on $C_2$ could come either from a shortcoming of the model or from an undetected systematic error in the experimental data. The importance of $C_2$ is largest at highest tensions, as it is $C_2$ which gives the curve its concavity, and it should be noted that the maximum applied tension $\sigma$ is around $4.5 \,\mathrm{mN/m}$ in Ref.~\cite{Goulian98}, which is comparable to the rupture threshold of $3-10 \,\mathrm{mN/m}$~\cite{Rawicz00}. The membrane properties may be affected at such high tensions in a way that is no longer well described by standard elastic models. It would be interesting to have more experimental data on the behavior of gramicidin channels under tension to see if this unexpected value of $C_2$ persists.

\begin{figure}[htb]
\centerline{\includegraphics[width=0.6\textwidth]{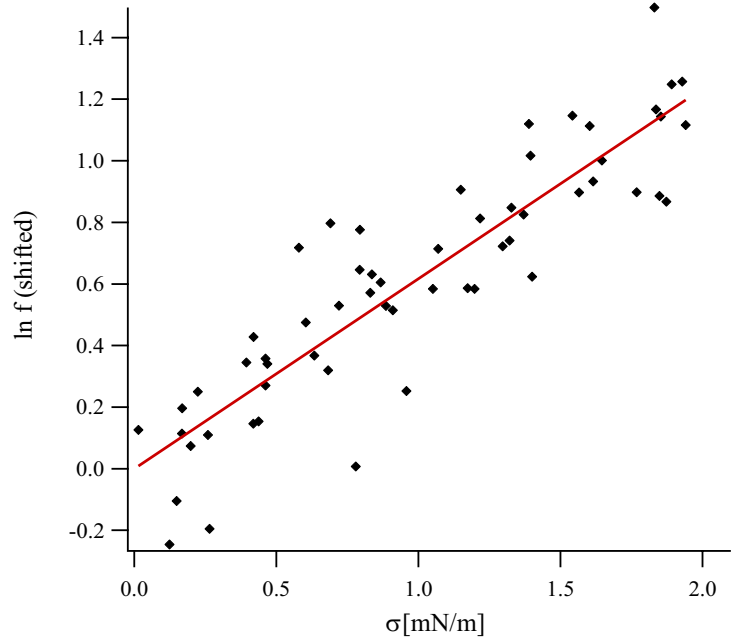}}
\caption{\textbf{Formation rate $f$ of gramicidin channels as a function of the applied tension $\sigma$, analyzed with a linear model, for $\sigma<2\,\mathrm{mN/m}$.} Diamonds: experimental data retrieved from Fig.~6b of Ref.~\cite{Goulian98}, after subtraction of the baselines $C_0^k$ (which are different from those of Fig.~\ref{fig:quadratic_fit} since the fitting model is here linear instead of quadratic). Line: best linear fit, yielding $C_1 = (0.62\pm0.05) \,\mathrm{(mN/m)}^{-1}$; correlation coefficient: $r=0.894$.}
\label{fig:linear_fit}
\end{figure}

Following the hypothesis that high tensions are problematic, we performed a linear fit of the data of Ref.~\cite{Goulian98} (i.e., a fit with $C_2=0$), keeping only the points corresponding to $\sigma<2\,\mathrm{mN/m}$: this yields $C_1 = (0.62\pm0.05) \,\mathrm{(mN/m)}^{-1}$ (see Fig.~\ref{fig:linear_fit}). In Table~\ref{tableC1C2}, we list, for different boundary conditions, the value of $k'_a$ which gives $C_1=0.62 \,\mathrm{(mN/m)}^{-1}$, and the value of $C_2$ obtained from our model for this $k'_a$. These values correspond to those that give the best agreement between our model and the linear fit to the low-tension data of Ref.~\cite{Goulian98} presented in Fig.~\ref{fig:linear_fit}. Table~\ref{tableC1C2} shows that the values of $k'_a$ that yield the best agreement with the experimental data have a similar order of magnitude as those obtained above with the full quadratic fit (see Table~\ref{tableC1C2_0}), remaining below $100\,\mathrm{mN/m}$. Again, these values depend a lot on $s$ for fixed-slope boundary-conditions. (For instance, the slope $s=-0.17$ is consistent with $k'_a=0$ (see Table~\ref{tableC1C2}). However, there is no \emph{a priori} reason for assuming that $k'_a=0$.) 

Again, we may compare our estimates of $k'_a$ (see Tables~\ref{tableC1C2_0} and~\ref{tableC1C2}) to the term $-\kappa_0(c_0-c'_0\Sigma_0)/d_0$, which also contributes to $K'_a$: here, $-\kappa_0(c_0-c'_0\Sigma_0)/d_0=-0.76\,\mathrm{mN/m}$. This is much smaller in absolute value than the corresponding estimates obtained from the numerical data of Ref.~\cite{West09} and of Ref.~\cite{Brannigan07}: here, as in the membranes studied in Ref.~\cite{Elliott83}, the neutral surface of a monolayer and its hydrophilic-hydrophobic interface are very close, while $\xi$ seemed to be of a few \AA~in the numerical simulations. We note in passing that this hints at a relevant difference between simulated membranes and real membranes. Besides, in the case of the free-slope boundary condition, our results imply that $k'_a$ should be the dominant contribution to $K'_a$ for the membranes studied in Ref.~\cite{Goulian98}, as for those of Ref.~\cite{Elliott83}.

Hence, for the free-slope boundary condition, our analyses of the numerical data of Ref.~\cite{West09} and of Ref.~\cite{Brannigan07}, and our analyses of the experimental data of Ref.~\cite{Elliott83} and of Ref.~\cite{Goulian98} all converge toward a value of a few tens of mN/m for $k'_a$, which is of the order of magnitude expected if $k'_a=\gamma/2$. Conversely, for the fixed-slope boundary condition, the value of $k'_a$ is coupled to that of the slope $s$.

\section*{Conclusion}
We have put forward a modification of membrane elastic models used to describe thickness deformations at the nanoscale. We have shown that terms involving the gradient (and the Laplacian) of the area per lipid contribute to important terms of the effective Hamiltonian of the bilayer membrane. We have reanalyzed numerical and experimental data to find some signature of the presence of these terms. Using the free-slope boundary condition at the boundary of the mismatched protein, we have obtained consistent results showing that the term stemming from the gradient of the area per molecule has a prefactor $k'_a$ in the range $13-60\,\mathrm{mN/m}$. Such values are consistent with the idea that this term involves a significant contribution of the interfacial tension $\gamma$ between water and the hydrocarbon-like hydrophobic part of the membrane. Indeed, this contribution should yield $k'_a=\gamma/2\approx25\,\mathrm{mN/m}$.  

Interestingly, our analysis of the experimental data from Ref.~\cite{Goulian98} has shown that these nice experimental results were not as well understood as assumed in the literature. Hence, it would be interesting to have more data on the behavior of gramicidin channels in membranes under tension.

Finally, the effective linear spring model~\cite{Nielsen98,Lundbaek99} is a very useful simplification of membrane elastic models when dealing with local thickness deformations and hydrophobic mismatch. Its applicability has been thoroughly tested on systems where gramicidin is used to probe the influence of various molecules on membrane properties (see, e.g., Ref.~\cite{Lundbaek10}). As other quadratic elastic models, our model yields an effective spring model. However, since the expression of the spring constant depends on the details of the model, careful consideration is required when one is interested in the behavior of a particular material constant.

\section*{Methods}
\subsection{Derivation of the effective Hamiltonian}
\label{AppA}
\subsubsection{General expression of the bilayer effective Hamiltonian}
\label{Deriv}
Let us consider a patch of bilayer membrane with a fixed projected area $A_p$, at fixed chemical potential $\mu$. The rest of the membrane (e.g., of the vesicle) plays the part of the reservoir that sets the chemical potential $\mu$. The effective Hamiltonian per unit projected area in each monolayer is $f^\pm=f_m^\pm/\bar\Sigma^\pm$, where $f_m^\pm$ is given by Eq.~\ref{fmolec}, while the projected area $\bar\Sigma^\pm$ per molecule reads $\bar \Sigma^\pm=\Sigma^\pm[1-(\mathbf{\nabla}h^\pm)^2/2]$ to second order. Hence, Eq.~\ref{fmolec} yields, to second order in the deformation and in the relative stretching of the monolayers,
\begin{align}
f^\pm&=-\frac{\mu}{\bar\Sigma^\pm}+\frac{f''_0}{2}\,\Sigma_0\,\left(\frac{\Sigma^\pm-\Sigma_0}{\Sigma^\pm}\right)^2\pm \frac{f_1}{2}\,\frac{\nabla^2 h^\pm}{\bar\Sigma^\pm} \pm \frac{f'_1}{2}\left(\frac{\Sigma^\pm-\Sigma_0}{\Sigma^\pm}\right)\nabla^2 h^\pm\nonumber\\
&+\frac{f_2}{4}\,\frac{\left(\nabla^2 h^\pm\right)^2}{\Sigma_0}+f_K\,\frac{\det(\partial_i\partial_j h^\pm)}{\Sigma_0} +\alpha\,\frac{(\mathbf{\nabla}\Sigma^\pm)^2}{\Sigma_0}+\beta\,\frac{\nabla^2\Sigma^\pm}{\bar\Sigma^\pm}+\zeta\,\frac{(\nabla^2\Sigma^\pm)^2}{\Sigma_0}\,.
\label{freloaded}
\end{align}

We assume that the hydrophobic chains of the lipids are incompressible. Let us introduce the excess hydrophobic thickness $u^+$ (resp. $u^-$) of the upper (resp. lower) monolayer, defined as its hydrophobic thickness along the normal to its hydrophilic-hydrophobic interface minus the equilibrium monolayer hydrophobic thickness $d_0/2$ (see Fig.~\ref{Dessin_defs}). In the spirit of Refs.~\cite{Dan93, ArandaEspinoza96, Brannigan06}, we use the incompressibility condition
\begin{equation}
 v=\Sigma^\pm\,\left(u^\pm+\frac{d_0}{2}\right)\,, \label{incomp}
\end{equation}
where $v$ is the constant hydrophobic volume per lipid. (In this incompressibility condition, a correction arising from membrane curvature is neglected. Using the complete incompressibility condition instead of this one yields the same effective Hamiltonian Eq.~\ref{jbarc2b}, but with different expressions of $c_0$ and $\kappa$ as a function of the constants involved in Eq.~\ref{fmolec}. These expressions depend on $\mu$, and consequently on the applied tension, but this dependence is negligible for realistic tension values. As the rest of our discussion is not affected by this, we keep the approximate incompressibility condition for the sake of simplicity. Note that the exact incompressibility condition was implemented recently in Ref.~\cite{Watson11}.)

In all the following, we will work to second order in the small dimensionless variables $u^\pm/d_0$, $|\mathbf{\nabla} u^\pm|$, $d_0\nabla^2 u^\pm$, $|\mathbf{\nabla} h^\pm|$ and $d_0\nabla^2 h^\pm$. In this framework, using the relations $\left(\Sigma_0-\Sigma^\pm\right)/\Sigma^\pm=2\,u^\pm/d_0$ and $u^\pm\nabla^2 u^\pm=\mathbf{\nabla}\left( u^\pm\mathbf{\nabla}u^\pm\right)-(\mathbf{\nabla}u^\pm)^2$, Eq.~\ref{freloaded} becomes
\begin{align}
f^\pm&=-\frac{\mu}{\Sigma_0}\left(1+\frac{2\,u^\pm}{d_0}+\frac{(\mathbf{\nabla}h^\pm)^2}{2}\right)+\frac{K_a}{d_0^2}(u^\pm)^2\pm\frac{\kappa_0\,c_0}{2}\nabla^2 h^\pm\pm\frac{\kappa_0}{d_0}(c_0-c'_0\Sigma_0)\,u^\pm\,\nabla^2 h^\pm \nonumber\\
&+\frac{\kappa_0}{4}(\nabla^2 h^\pm)^2 +\frac{\bar\kappa}{2}\det(\partial_i\partial_j h^\pm)\nonumber\\
&+k'_a\left(\mathbf{\nabla}u^\pm\right)^2+\frac{2\,\beta}{d_0}\left[\frac{2}{d_0}\mathbf{\nabla}\cdot(u^\pm\mathbf{\nabla}u^\pm)-\nabla^2 u^\pm\right]+k''_a\,d_0^2(\nabla^2 u^\pm)^2\,.\label{onemore}
\end{align}
In this expression, we have introduced the constitutive constants of a monolayer: $f''_0\Sigma_0=K_a/2$ is compressibility modulus of the monolayer, $f_2/(2\Sigma_0)=\kappa_0/2$ is its bending rigidity, $f_K/\Sigma_0=\bar\kappa/2$ is its Gaussian bending rigidity, $f_1/f_2=c_0$ is its spontaneous (total) curvature, and $f'_1/f_2=c'_0$ is the modification of the spontaneous (total) curvature due to area variations. More precisely, $c'_0=dc_s/d\Sigma$ where $c_s(\Sigma)=c_0+c'_0(\Sigma-\Sigma_0)$ is the lipid area-dependent (total) spontaneous curvature of the monolayer. In addition, recall that $d_0$ denotes the equilibrium hydrophobic thickness of the bilayer membrane. Finally, we have introduced the constants \begin{align}
k'_a&=4\,\frac{\alpha\,\Sigma_0+\beta}{d_0^2}\,,\label{kpa}\\
k''_a&=4\,\frac{\zeta\,\Sigma_0}{d_0^4}\,.\label{ksa}                                                                                                                                                                                                                                                                                                                                                                                                                                                                                                                                                                                                                                                                                                                                                                                                                                                                                                                                                                                                                                                                                                                                                                                                                                                                                           \end{align}
These two constants have the dimension of a surface tension, like $K_a$.

In our description, the state of monolayer $\pm$ is determined by the two variables $h^\pm$ and $u^\pm$. Hence, the state of the bilayer membrane is \emph{a priori} determined by four variables. However, given that there must be no space between the two monolayers, the distance along $z$ between the hydrophilic-hydrophobic interfaces of the two monolayers must be equal to the sum of their projected thicknesses. Hence, to second order, we have the following geometrical constraint: 
\begin{equation}
h^+-h^-=\left(u^++\frac{d_0}{2}\right)\left[1-\frac{(\mathbf{\nabla}h^+)^2}{2}\right]+\left(u^-+\frac{d_0}{2}\right)\left[1-\frac{(\mathbf{\nabla}h^-)^2}{2}\right]\,.\label{ctrg}
\end{equation}
This leaves us with only three independent variables to describe the state of the membrane. Let us choose the average shape $h$ of the bilayer, the sum $u$ of the excess hydrophobic thicknesses of the two monolayers, and the difference $\delta$ between them:
\begin{align}
h&=\frac{h^++h^-}{2}\,,\label{def_var}\\
u&=u^++u^-\,,\\
\delta &=u^+-u^-\,.
\end{align} 

Thus, we can rewrite the effective Hamiltonian $f=f^++f^-$ per unit projected area of the membrane in terms of the new variables $h$, $u$ and $\delta$. It reads, to second order in the small dimensionless variables $u/d_0$, $\delta/d_0$, $|\mathbf{\nabla} u|$, $|\mathbf{\nabla} \delta|$, $|\mathbf{\nabla} h|$, $d_0\nabla^2 u$, $d_0\nabla^2 \delta$, and $d_0\nabla^2 h$, and discarding derivatives of order higher than two: 
\begin{align}
f&=\sigma\left[1+\frac{u}{d_0}+\frac{(\mathbf{\nabla}h)^2}{2}+\frac{(\mathbf{\nabla}u)^2}{8}\right]+\frac{K_a}{2\,d_0^2}\left(u^2+\delta^2\right) +\frac{\kappa_0}{2}\left[(\nabla^2 h)^2+\frac{1}{4}(\nabla^2 u)^2\right]\nonumber\\
&+\frac{\kappa_0 \,c_0}{2}\,\nabla^2 u+\frac{\kappa_0}{2\,d_0}(c_0-c'_0\Sigma_0)\left(u\,\nabla^2 u+2\,\delta\,\nabla^2 h\right)\nonumber\\ &+\bar\kappa\left[\det(\partial_i\partial_j h) +\frac{1}{4}\det(\partial_i\partial_j u)\right]\nonumber\\
&+\frac{k'_a}{2}\left[(\mathbf{\nabla}u)^2+(\mathbf{\nabla}\delta)^2\right]+\frac{k''_a\,d_0^2}{2}\left[(\nabla^2 u)^2+(\nabla^2 \delta)^2\right]\nonumber\\
&-\frac{2\,\beta}{d_0}\,\nabla^2 u+\frac{2\,\beta}{d_0^2}\left[\,\mathbf{\nabla}\cdot(u\mathbf{\nabla}u)+\mathbf{\nabla}\cdot(\delta\mathbf{\nabla}\delta)\,\right]\,,
\label{jbar}
\end{align}
where we have introduced $\sigma=-2\mu/\Sigma_0$, which plays the part of an externally applied tension (see Methods, Sec.~\ref{Tension}).

\subsubsection{Eliminating $\delta$} 
\label{elim_deltat}
In the present study, we are not interested in the variable $\delta$. In a coarse-graining procedure, this degree of freedom can be eliminated by integrating over it. In our Gaussian theory, it simply amounts to minimizing $f$ with respect to $\delta$. This variable is coupled to the membrane curvature $\nabla^2 h$, but not to $u$. In the case of a constant curvature, the constant value
\begin{equation}
\delta=-\frac{d_0\,\kappa_0}{K_a}(c_0-c'_0\Sigma_0)\,\,\nabla^2 h\,,
\end{equation}
is a simple solution to the Euler-Lagrange equations in $\delta$, for which the term involving $\delta$ in $f$ reads
\begin{align}
f_{\delta}=-\frac{1}{2}\,\frac{\kappa_0^2}{K_a}(c_0-c'_0\Sigma_0)^2\,\,(\nabla^2 h)^2\,.
\end{align}
As the variable $\delta$ varies spontaneously on length scales much shorter than the variable $h$, we can consider in a first approximation that $\delta$ will simply follow $\nabla^2 h$, in which case this constant solution is the valid one. Thus, after this partial minimization, this term provides a correction to $\kappa_0$.

We finally obtain
\begin{align}
 f&=\sigma\left[1+\frac{u}{d_0}+\frac{(\mathbf{\nabla}h)^2}{2}\right]+\frac{K_a}{2\,d_0^2}\,u^2+\left[\frac{k'_a}{2}+\frac{\sigma}{8}-\frac{\kappa_0}{2\,d_0}(c_0-c'_0\Sigma_0)\right](\mathbf{\nabla}u)^2\nonumber\\
& +\frac{\kappa}{2}(\nabla^2 h)^2+\left(\frac{\kappa_0}{8}+\frac{k''_a\,d_0^2}{2}\right)(\nabla^2 u)^2+\bar\kappa\left[\det(\partial_i\partial_j h) +\frac{1}{4}\det(\partial_i\partial_j u)\right]\nonumber\\
&+\left[\frac{\kappa_0 \,c_0}{2}-\frac{2\,\beta}{d_0}\right]\,\nabla^2 u+\left[\frac{\kappa_0}{2\,d_0}(c_0-c'_0\Sigma_0)+\frac{2\,\beta}{d_0^2}\right]\,\mathbf{\nabla}\cdot(u\mathbf{\nabla}u) \,,
\label{moregenbis}
\end{align}
where the usual Helfrich bending rigidity $\kappa$, associated with the average shape, is related to $\kappa_0$ through
\begin{equation}
\kappa=\kappa_0 -\frac{\kappa_0^2}{K_a}(c_0-c'_0\Sigma_0)^2\,.
\label{kappakappa0}
\end{equation}

In the case where the average shape of the membrane is flat, i.e., $h=0$, dropping constant terms, we obtain the expression of $f$ in Eq.~\ref{jbarc2b} with
\begin{align}
K'_a&=-\frac{\kappa_0}{d_0}(c_0-c'_0\Sigma_0)+k'_a+\frac{\sigma}{4}\,,\\
K''_a&=\frac{\kappa_0}{4}+k''_a d_0^2\,,\label{K''a_gen}\\
A_1&=\frac{\kappa_0 \,c_0}{2}-\frac{2\,\beta}{d_0}\,,\\
A_2&=\frac{\kappa_0}{2\,d_0}(c_0-c'_0\Sigma_0)+\frac{2\,\beta}{d_0^2}\,.
\label{A1A2beta}
\end{align}

Thus, in general, in Eq.~\ref{jbarc2b}, the constants $K'_a$, $K''_a$ include contributions in $k'_a$ and $k''_a$, which arise from $\alpha$, $\beta$ and $\zeta$ (see Eqs.~\ref{kpa},~\ref{ksa}). Therefore, the terms in gradient and Laplacian of $\Sigma$ introduced in Eq.~\ref{fmolec} cannot be neglected \emph{a priori}, as they contribute to the terms in $(\mathbf{\nabla}u)^2$ and $(\nabla^2 u)^2$ that are traditionally accounted for in models describing membrane thickness deformations~\cite{Huang86, Dan93, ArandaEspinoza96, Goulian98, Nielsen98, Brannigan06}. Due to these contributions, the values of the constants $K'_a$ and $K''_a$ are not fully predicted by the constants involved in the Helfrich model. This stands in contrast with the models developed previously~\cite{Huang86, Dan93, ArandaEspinoza96, Goulian98, Nielsen98, Brannigan06}. In addition, the terms arising from $\alpha$, $\beta$ and $\zeta$ modify the relations between the various coefficients: in the previous models that accounted for boundary terms, assuming $\alpha=\beta=\zeta=0$, and disregarding tension, one had $K'_a=-2\,A_2$~\cite{Brannigan06}, which is no longer true here. This will affect the equilibrium thickness profile of a membrane containing a mismatched protein.

\subsubsection{Link with the Helfrich Hamiltonian}
\label{special_cases}
Since the variables $h$ and $u$ are decoupled in the Hamiltonian density $f$ given by Eq.~\ref{moregenbis}, the terms depending on $h$ can be isolated, yielding
\begin{align}
f_h=\sigma\,\left[1+\frac{(\mathbf{\nabla}h)^2}{2}\right]+\frac{\kappa}{2}\,(\nabla^2 h)^2+\bar\kappa\det(\partial_i\partial_j h)\,,
\label{jbarc1}
\end{align}
which corresponds to the Helfrich Hamiltonian~\cite{Helfrich73} for a membrane composed of two identical monolayers. In particular, the term in $\sigma$ has the standard form of a Helfrich tension term, conjugate to the actual area $A$ of the membrane, since the element of area is $dA=dxdy\sqrt{1+(\mathbf{\nabla}h)^2}=dxdy\left[1+(\mathbf{\nabla}h)^2/2\right]$ to second order. Hence, $\sigma$ can be viewed as an effective applied tension. This interpretation of $\sigma$ is explained in more detail in Sec.~\ref{Tension} of our Methods part. 

Hence, our model gives back the Helfrich Hamiltonian if the state of the membrane is described only by its average shape $h$, i.e., if the variable $u$ is integrated out. 

\subsubsection{Stability criterion}
\label{Stab}
Let us focus on a membrane with flat average shape $h$, described by Eq.~\ref{jbarc2b}. Depending on the values of the constants $K_a$, $K'_a$ and $K''_a$, a homogeneous thickness $u=0$ can be less or more energetically favorable than an undulated shape. The physical situation we wish to describe is the one where the equilibrium state has a homogeneous thickness. To determine which sets of constants comply with this, let us calculate the effective Hamiltonian per unit projected area $f_\mathrm{def}$ of a membrane with harmonic undulations characterized by the wave vector $q$. Neglecting boundary terms (by taking appropriate boundary conditions or by assuming that the undulations decay on some large length scale), we obtain $f_\mathrm{def}\propto K_a/d_0^2+K'_a q^2+K''_a q^4$, where the omitted prefactor is positive. 
The flat shape is favored if $f_\mathrm{def}>0$ for all $q$, and otherwise there exist some values of $q$ for which it is unstable. Thus, the conditions for the stability of the flat shape are $K_a>0$, $K''_a>0$ and $K'_a>-2\sqrt{K_a K''_a}/d_0$.

\subsection{Membrane submitted to an external tension}
\label{Tension}
In Sec.~\ref{AppA} of our Methods part, we have derived the effective Hamiltonian of a bilayer membrane in the $(\mu, A_p)$ ensemble. This is the most convenient thermodynamic ensemble to work in. However, in order to describe experiments where a vesicle is submitted to an external tension, one should work in the $(N, \tau)$ ensemble, where $N$ is the number of lipids in the vesicle and $\tau$ is the externally applied tension. This is especially interesting in order to analyze the results of Ref.~\cite{Goulian98}. The ensemble change can be performed using a Legendre transformation: in the $(N, \tau)$ ensemble, the adapted effective Hamiltonian is $G(N, \tau)=F(\mu, A_p)+\mu N-\tau A_p$, where $F(\mu, A_p)=\int_{A_p}dxdy\, f$, with $f$ expressed in Eq.~\ref{moregenbis}, and 
\begin{equation}
N=-\left.\frac{\partial F}{\partial \mu}\right|_{A_p},\,\,\,\tau=\left.\frac{\partial F}{\partial A_p}\right|_{\mu}\,.\label{Ntau}
\end{equation}

Let us restrict ourselves to the case of a homogeneous and flat membrane, i.e., to a membrane with constant $h$ and $u$. Then, using Eq.~\ref{Ntau} to eliminate the variables $\mu$ and $A_p$ from the expression of $G$, we obtain, to second order:
\begin{equation}
G(N,\tau)=N\,\frac{v}{d_0}\left[-\tau+\tau\frac{u}{d_0}+\left(K_a-2\,\tau\right)\frac{u^2}{2\,d_0^2}\right]\,.
\end{equation}

Minimizing $G$ with respect to $u$ yields the equilibrium excess thickness $u_\mathrm{eq}$ of the membrane at a given imposed tension $\tau$. To first order, it reads
\begin{equation}
u_\mathrm{eq}=-\frac{\tau}{K_a}\,d_0\,,
\label{ueq1}
\end{equation}
Note that, since $u/d_0$ is assumed to be a first-order quantity, $\tau/K_a$ must be first-order too for our description to be valid for $u=u_\mathrm{eq}$. This property has been used to simplify the result in Eq.~\ref{ueq1}. In practice, $\tau\ll K_a$ is well verified, given that $\tau$ cannot exceed a few mN/m without the vesicle bursting, while $K_a$ is of order $100\,\mathrm{mN/m}$.
Since $d_0$ is the equilibrium hydrophobic thickness of this piece of homogeneous and flat membrane submitted to a vanishing external tension, it is consistent that $u_\mathrm{eq}$ vanishes when $\tau$ does, as $u$ is the excess thickness with respect to $d_0$. Eq.~\ref{ueq1} shows that the thickness of a membrane with fixed number of lipids decreases when the external tension increases, and is in agreement with Ref.~\cite{Goulian98}.

We are now going to show that the constant $\sigma$ in the $(\mu, A_p)$ ensemble (see, e.g., Eq.~\ref{moregenbis}) plays the part of an externally applied tension. For this, let us calculate the equilibrium thickness of a membrane patch with projected area $A_p$ at a chemical potential $\mu$, when it is homogeneous and flat. This amounts to minimizing $f$ with respect to $u$. For a homogeneous and flat membrane, Eq.~\ref{moregenbis} becomes
\begin{equation}
f=\sigma\left(1+\frac{u}{d_0}\right)+\frac{K_a}{2}\frac{u^2}{d_0^2}\,,
\end{equation}
Minimizing $f$ with respect to $u$ then gives
\begin{equation}
u_\mathrm{eq}=-\frac{\sigma}{K_a}\,d_0\,.
\label{deq}
\end{equation}
Comparing Eq.~\ref{deq} to Eq.~\ref{ueq1} shows that $\sigma$ plays the part of the externally applied tension $\tau$. Hence, $\sigma$ can be considered as an effective applied tension.

\subsection{Membrane containing a cylindrical mismatched protein}
In this Section, we write down explicitly the equilibrium shape and the deformation energy of a membrane which contains a single cylindrical transmembrane protein with a hydrophobic mismatch (see Fig.~\ref{Dessin_defs}B). This protein can correspond to a gramicidin channel in the dimer state. We focus on a membrane with a flat average shape, described by the effective Hamiltonian per unit projected area in Eq.~\ref{jbarc2b}. We denote the radius of the protein by $r_0$, and its hydrophobic thickness by $\ell$. We take the center of the cylindrical protein as the origin of the frame, which yields cylindrical symmetry.

In order to treat the case where the membrane is submitted to a tension $\sigma$, we rewrite Eq.~\ref{jbarc2b} in terms of the variable $\tilde u=u-u_\mathrm{eq}=u+\sigma d_0/K_a$, which represents the excess hydrophobic thickness of the bilayer relative to its equilibrium value at an applied tension $\sigma$ (see Eq.~\ref{deq}). Discarding constant terms and using the relation $\sigma\ll K_a$, which yields $\sigma d_0 A_2/K_a\ll A_1$, it yields
\begin{align}
f&=\frac{K_a}{2\,d_0^2}\,\tilde u^2+\frac{K'_a}{2}\,(\mathbf{\nabla}\tilde u)^2+\frac{K''_a}{2}\,(\nabla^2 \tilde u)^2\nonumber\\
&+A_1\,\nabla^2 \tilde u+A_2\,\mathbf{\nabla}\cdot(\tilde u\mathbf{\nabla}\tilde u) +\frac{\bar\kappa}{4}\,\det(\partial_i\partial_j \tilde u)\,.
\label{jbarc2bb}
\end{align}

\subsubsection{Equilibrium thickness profile}
\label{ThickProf}
Let us first review (see, e.g., Ref.~\cite{Nielsen98}) the equilibrium thickness profile $\tilde u$ of the membrane containing the mismatched protein. This equilibrium shape is solution to the Euler-Lagrange equation associated with the effective Hamiltonian in Eq.~\ref{jbarc2bb},
\begin{equation}
\nabla^4 \tilde u- \frac{K'_a}{K''_a} \nabla^2\tilde u+\frac{K_a}{K''_a d_0^2}\tilde u=0\,.
\label{EL}
\end{equation}

Using the cylindrical symmetry of the problem and choosing solutions that vanish at infinity, we obtain, if the stability condition Eq.~\ref{stab} is verified, the following solution to the Euler-Lagrange equation Eq.~\ref{EL}:
\begin{equation}
\tilde u (r)=A_+ \mathrm{K}_0(k_+ r)+A_- \mathrm{K}_0(k_- r)\,,
\label{sol_fix}
\end{equation}
where $\mathrm{K}_n$ is the $n^\mathrm{th}$-order modified Bessel function of the second kind, and
\begin{align}
k_\pm=\frac{1}{\sqrt{2}}\left\{\frac{K'_a}{K''_a}\pm\left[\left(\frac{K'_a}{K''_a}\right)^2-4\frac{K_a}{K''_a d_0^2}\right]^{1/2}\right\}^{1/2}\,,
\label{kpm}
\end{align}
which are either both real or complex conjugate.

The integration constants $A_\pm$ are determined by the boundary conditions at $r=r_0$. The first boundary condition corresponds to strong hydrophobic coupling: on the inclusion boundary, the hydrophobic thickness of the membrane is equal to that of the inclusion, which is denoted by $\ell$ (see Fig.~\ref{Dessin_defs}B). It yields $u(r_0)=u_0=\ell-d_0$ (to first order, as explained in our Section entitled ``Deformation profiles close to a mismatched protein''), or equivalently $\tilde u (r_0)=\tilde u_0=\ell-d_0\left(1-\sigma/K_a\right)$. As far as the second boundary condition at $r=r_0$ is concerned, we will treat explicitly two different cases, which correspond respectively to a fixed slope and to a free slope in $r_0$, as explained in the main text of the article. \\

\textbf{Fixed slope.} In the case where the boundary conditions in $r=r_0$ are
\begin{equation}
\left\{ \begin{array}{l}
\tilde u (r_0)=\tilde u_0=\ell-d_0\left(1-\displaystyle{\frac{\sigma}{K_a}}\right)\\
\tilde u' (r_0)=s
\end{array} \right. \,,
\label{pentefixe}
\end{equation}
which corresponds to a strong hydrophobic coupling and a fixed slope $s$ at $r=r_0$, we obtain:
\begin{equation}
A_\pm=\frac{\mathrm{K}_0^\mp s+ k_\mp \mathrm{K}_1^\mp \tilde u_0}{ k_\mp \mathrm{K}_0^\pm \mathrm{K}_1^\mp - k_\pm \mathrm{K}_0^\mp \mathrm{K}_1^\pm}\,,
\label{Apm}
\end{equation}
where
\begin{equation}
\mathrm{K}_n^\pm=\mathrm{K}_n\left(k_\pm r_0\right)\,.
\label{defKn}
\end{equation}
Note that $A_+$ and $A_-$ are either both real or complex conjugate (like $k_\pm$), which ensures that the solution Eq.~\ref{sol_fix} is real. \\

\textbf{Free slope. }An alternative choice of boundary conditions in $r=r_0$ is
\begin{equation}
\left\{ \begin{array}{l}
\tilde u (r_0)=\tilde u_0=\ell-d_0\left(1-\displaystyle{\frac{\sigma}{K_a}}\right)\\
\left.\left(K''_a \nabla^2 \tilde u+\displaystyle{\frac{\bar\kappa}{4}\frac{\tilde u'}{r}}+A_2\tilde u+A_1\right)\right|_{r=r_0}=0
\end{array} \right. \,,
\label{pentelibre}
\end{equation}
to first order again. The first of these conditions corresponds to a strong hydrophobic coupling, as before. The second one arises from minimizing the total free energy of the system without further constraints. It corresponds to the case where the slope at $r=r_0$ is free to adjust itself to yield the smallest deformation energy. With these ``free-slope'' boundary conditions, we obtain:
\begin{equation}
A_\pm=\pm
\frac{ \bar\kappa\, k_\mp \mathrm{K}_1^\mp \tilde u_0-4 r_0 \mathrm{K}_0^\mp \left[A_1+\left(A_2+K''_a
   k_\mp^2\right)\tilde u_0 \right]}{4 r_0 K''_a \left(k_+^2-k_-^2\right) \mathrm{K}_0^+ \mathrm{K}_0^- - \bar\kappa\left( k_+
   \mathrm{K}_0^- \mathrm{K}_1^+ - k_- \mathrm{K}_0^+ \mathrm{K}_1^-\right)}\,,
\label{Apm_b}
\end{equation}
which are, again, either both real or complex conjugate.

Let us now assume that $\beta=\zeta=0$, as in the main text of this article. In order to understand the impact of $k'_a$ (i.e., of $\alpha$) on $A_\pm$ in the free-slope case, let us express $A_\pm$ as a function of $k'_a$, $r_0$, $d_0$ and of the bulk constants $K_a$, $K'_a$ and $K''_a$, whose values can be extracted from the fluctuation spectra in simulations. Using Eq.~\ref{kpm}, the relation $A_1=2K''_a c_0$, which can be derived from Eqs.~\ref{A2} and~\ref{Ksa}, and the relation $A_2=(k'_a-K'_a)/2$, which stems from Eqs.~\ref{Kpa} and~\ref{A2}, we obtain:
\begin{equation}
A_\pm=\pm
\frac{ \bar\kappa\, k_\mp \mathrm{K}_1^\mp \tilde u_0-2 r_0 \mathrm{K}_0^\mp \left\{4K''_a c_0+\left[k'_a\pm K''_a\left(
   k_-^2-k_+^2\right)\right]\tilde u_0 \right\}}{4 r_0 K''_a \left(k_+^2-k_-^2\right) \mathrm{K}_0^+ \mathrm{K}_0^- - \bar\kappa\left( k_+
   \mathrm{K}_0^- \mathrm{K}_1^+ - k_- \mathrm{K}_0^+ \mathrm{K}_1^-\right)}\,.
\label{Apm_b_c0}
\end{equation}
For fixed values of $r_0$, $d_0$, $K_a$, $K'_a$ and $K''_a$, the constants $A_\pm$ can be viewed simply as functions of $k'_a$ and $c_0$: let us denote them by $A_\pm(k'_a,c_0)$. The following relation holds for all $k'_a$ and $c_0$:
\begin{equation}
A_\pm\left(k'_a,c_0\right)=A_\pm\left(0,\tilde c_0\right)\,,
\label{corresp}
\end{equation}
with 
\begin{equation}
\tilde c_0=c_0+\frac{k'_a}{4K''_a}\tilde u_0\,.
\end{equation}
Hence, in the framework of a model that assumes $k'_a=0$, the effect of a nonvanishing $k'_a$ on the equilibrium membrane thickness profile would be that $c_0$ is replaced by a renormalized spontaneous curvature $\tilde c_0$, which depends linearly on $\tilde u_0$. At vanishing applied tension (in which case, $\tilde u_0=u_0$), and neglecting the difference between $\kappa_0=K''_a/4$ and $\kappa$, we obtain Eq.~\ref{c0u0}.

\subsubsection{Deformation energy}
\label{defen}
Let us now calculate the deformation energy $F$ of the membrane due to the presence of the mismatched protein.
For the equilibrium shape of the membrane, which is solution to the Euler-Lagrange equation Eq.~\ref{EL}, we are left only with boundary terms at the inclusion edge in $r=r_0$ (no other boundary terms contribute, since the deformation $\tilde u$ caused by the presence of the mismatched channel vanishes sufficiently far away from it). We can write
\begin{align}
F&=\int_{A_p}dxdy\,f=2\pi\int_{r_0}^\infty rdr\,f\nonumber\\
&=\pi \Bigg\{K''_a \,r\left[\tilde u\,\frac{d}{dr}\left(\nabla^2\tilde u\right)-\tilde u'\,\nabla^2\tilde u-\frac{K'_a}{K''_a}\,\tilde u\,\tilde u'\right]\nonumber\\
&-2\left[A_1 \,r \,\tilde u'+A_2 \,r \,\tilde u \,\tilde u'+\frac{\bar\kappa}{8}\,\tilde u'^2\right]\Bigg\}\Bigg|_{r=r_0}\,,
\label{en_def}
\end{align}
where $\tilde u'= d \tilde u/dr$. We have used the expression of the Gaussian curvature for small deformations in a system with cylindrical symmetry: $\det(\partial_i\partial_j \tilde u)=\tilde u'\tilde u''/r=(2\,r)^{-1}\,d(\tilde u'^2)/dr$. To express the deformation energy $F$ explicitly, one has to use the boundary conditions in $r=r_0$.\\

\textbf{Fixed slope. } For the boundary conditions in Eq.~\ref{pentefixe}, corresponding to a fixed slope in $r_0$, using Eqs.~\ref{sol_fix},~\ref{kpm} and~\ref{Apm}, we can rewrite the deformation energy of the membrane in Eq.~\ref{en_def} as
\begin{align}
F&=-2\pi\left[A_1 \,r_0 \,s+A_2 \,r_0 \,\tilde u_0 \,s+\frac{\bar\kappa}{8}\,s^2\right]\nonumber\\
&+\frac{\pi\,r_0\,K''_a}{ k_+ \mathrm{K}_0^- \mathrm{K}_1^+ - k_-\mathrm{K}_0^+ \mathrm{K}_1^-} \Big[k_+ k_-\left(k_+^2-k_-^2\right) \mathrm{K}_1^+ \mathrm{K}_1^- \tilde u_0 ^2\nonumber\\
& +2\, k_+ k_-\left(k_+ \mathrm{K}_0^+ \mathrm{K}_1^- - k_- \mathrm{K}_0^- \mathrm{K}_1^+\right) \tilde u_0 \,s + \mathrm{K}_0^+ \mathrm{K}_0^-\left(k_+^2-k_-^2\right) s^2 \Big]\,.
\label{Fgen}
\end{align}
This expression shows that $F$ is a second-order polynomial in $\tilde u_0$ and $s$.

\textbf{Spring constant for $s=0$.} In the particular case where the fixed slope $s$ vanishes, Eq.~\ref{Fgen} becomes
\begin{equation}
F=H_0\tilde u_0^2\,,
\end{equation}
where the effective spring constant reads
\begin{equation}
H_0=\frac{\pi\,r_0\,K''_a\,k_+ k_-\left(k_+^2-k_-^2\right) \mathrm{K}_1^+ \mathrm{K}_1^-}{ k_+ \mathrm{K}_0^- \mathrm{K}_1^+- k_- \mathrm{K}_0^+ \mathrm{K}_1^-}\,.
\label{ressortfixe}
\end{equation}

\textbf{Dependence on applied tension.} Since $\tilde u_0=\ell-d_0(1-\sigma/K_a)$, Eq.~\ref{Fgen} shows that $F$ is a second-order polynomial in the applied tension $\sigma$. (In our model, $K'_a$ features a contribution coming from $\sigma$, see Eq.~\ref{Kpa}. However, as mentioned in the main text, the dependence of $K'_a$ on $\sigma$ is negligible in practice, and we thus disregard it: in this framework, $C_1$ and $C_2$ do not depend on $\sigma$.) We can write
\begin{equation}
-\frac{F}{k_\mathrm{B}T}=C_0+C_1 \sigma+C_2 \sigma^2\,,
\end{equation}
with
\begin{align}
C_1&= \frac{2 \pi  d_0 r_0\, K''_a\, k_+ k_-}{k_\mathrm{B}T\,K_a \left( k_- \mathrm{K}_0^+ \mathrm{K}_1^--k_+ \mathrm{K}_0^- \mathrm{K}_1^+ \right)}\Big[\left(k_+ \mathrm{K}_0^+ \mathrm{K}_1^- - k_- \mathrm{K}_0^- \mathrm{K}_1^+\right) s  \nonumber\\
&+\left(k_+^2-k_-^2\right)\mathrm{K}_1^- \mathrm{K}_1^+ \left(d_0-\ell\right)\Big]+\frac{2 \pi  d_0 r_0}{ k_\mathrm{B}T\,K_a} \,s A_2\,,\label{C1fixe}\\
C_2&=-\frac{d_0^2}{K_a^2}\,\frac{H_0}{k_\mathrm{B}T}\,, \label{C2fixe}
\end{align}
where $H_0$ is the effective spring constant expressed in Eq.~\ref{ressortfixe}. Note that $\bar \kappa$ and $A_1$ do not appear in the coefficients $C_1$ and $C_2$, and that $A_2$ and $s$ are only present in $C_1$.\\

\textbf{Free slope. }For the boundary conditions in Eq.~\ref{pentelibre}, corresponding to a free slope in $r_0$, using Eqs.~\ref{sol_fix},~\ref{kpm} and~\ref{Apm_b}, we can rewrite the deformation energy of the membrane (see Eq.~\ref{en_def}) as
\begin{align}
F&=\frac{\pi r_0}{ \bar\kappa\left( k_+ \mathrm{K}_0^- \mathrm{K}_1^+ - k_- \mathrm{K}_0^+ \mathrm{K}_1^-\right) - 4 r_0 K''_a \left(k_+^2-k_-^2\right) \mathrm{K}_0^+ \mathrm{K}_0^-} \times \nonumber\\
&\Bigg\{\Big[4 r_0 \Big( k_+ \mathrm{K}_0^- \mathrm{K}_1^+ \left(A_2 + K''_a k_-^2\right)^2 - k_- \mathrm{K}_0^+ \mathrm{K}_1^- \left(A_2 + K''_a k_+^2\right)^2 \Big) \nonumber\\
&+ K''_a \bar\kappa \left(k_+^2-k_-^2 \right) k_+ k_- \mathrm{K}_1^+ \mathrm{K}_1^-\Big]\tilde u_0^2  \nonumber\\
&+8 A_1 r_0 \Big[K''_a k_- k_+ \left( k_- \mathrm{K}_0^- \mathrm{K}_1^+ - k_+ \mathrm{K}_0^+ \mathrm{K}_1^-\right) \nonumber\\
& + A_2 \left( k_+ \mathrm{K}_0^- \mathrm{K}_1^+ - k_- \mathrm{K}_0^+ \mathrm{K}_1^-\right)\Big] \tilde u_0 +4 A_1^2 r_0\left(k_+ \mathrm{K}_0^- \mathrm{K}_1^+ - k_- \mathrm{K}_0^+ \mathrm{K}_1^- \right)\Bigg\}
\label{Fgen_b}
\end{align}
This expression shows that $F$ is a second-order polynomial in $\tilde u_0$.

\textbf{Spring constant.} Eq.~\ref{Fgen_b} can be expressed as
\begin{equation}
F=H_f\left(\tilde u_0-\tilde u_0^\mathrm{min}\right)^2+F^\mathrm{min}\,, \label{fhf}
\end{equation}
where the effective spring constant reads
\begin{align}
H_f&=\frac{\pi r_0}{ \bar\kappa\left( k_+ \mathrm{K}_0^- \mathrm{K}_1^+ - k_- \mathrm{K}_0^+ \mathrm{K}_1^-\right) - 4 r_0 K''_a \left(k_+^2-k_-^2\right) \mathrm{K}_0^+ \mathrm{K}_0^-} \times \nonumber\\
&\Big[4 r_0 \Big( k_+ \mathrm{K}_0^- \mathrm{K}_1^+ \left(A_2 + K''_a k_-^2\right)^2 - k_- \mathrm{K}_0^+ \mathrm{K}_1^- \left(A_2 + K''_a k_+^2\right)^2 \Big) \nonumber\\
&+ K''_a \bar\kappa \left(k_+^2-k_-^2 \right) k_+ k_- \mathrm{K}_1^+ \mathrm{K}_1^-\Big]\,,
\label{ressortlibre}
\end{align}
while $\tilde u_0^\mathrm{min}$ denotes the value of $\tilde u_0$ that minimizes $F$, and $F^\mathrm{min}$ is the minimum of $F$, obtained for $\tilde u_0=\tilde u_0^\mathrm{min}$. Note that both $u_0^\mathrm{min}$ and $F^\mathrm{min}$ are nonzero if $A_1\neq 0$ (see Eq.~\ref{Fgen_b}), due to the spontaneous curvature of each monolayer. The effect of monolayer spontaneous curvature was disregarded in Ref.~\cite{Nielsen98,Goulian98}, which explains why Eq.~\ref{fhf} differs from the standard expression $F=H_f u_0^2$~\cite{Nielsen98}. 

\textbf{Dependence on applied tension.} Since $\tilde u_0=\ell-d_0(1-\sigma/K_a)$, Eq.~\ref{Fgen_b} shows that $F$ is a second-order polynomial in the applied tension $\sigma$ (neglecting the $\sigma$-dependence of $K'_a$ as explained in the main text). Thus, we can write
\begin{equation}
-\frac{F}{k_\mathrm{B}T}=C_0+C_1 \sigma+C_2 \sigma^2\,,
\end{equation}
with
\begin{align}
C_1&=\frac{-2\pi r_0d_0}{k_\mathrm{B}T\,K_a \left[\bar\kappa\left( k_+ \mathrm{K}_0^- \mathrm{K}_1^+ - k_- \mathrm{K}_0^+ \mathrm{K}_1^-\right) - 4 r_0 K''_a \left(k_+^2-k_-^2\right) \mathrm{K}_0^+ \mathrm{K}_0^-\right]} \times \nonumber\\
 & \Bigg\{4 A_1 r_0 \Big[K''_a k_- k_+ \left( k_- \mathrm{K}_0^- \mathrm{K}_1^+ - k_+ \mathrm{K}_0^+ \mathrm{K}_1^- \right) \nonumber\\ 
      &   +A_2 \left( k_+ \mathrm{K}_0^- \mathrm{K}_1^+ - k_- \mathrm{K}_0^+ \mathrm{K}_1^-\right)\Big] + \Big[4 r_0 \Big( k_+ \mathrm{K}_0^- \mathrm{K}_1^+ \left(A_2 + K''_a k_-^2\right)^2  \nonumber\\ 
     & - k_- \mathrm{K}_0^+ \mathrm{K}_1^- \left(A_2 + K''_a k_+^2\right)^2\Big) +  K''_a \bar\kappa \left(k_+^2-k_-^2\right) k_+ k_- \mathrm{K}_1^+ \mathrm{K}_1^-\Big](\ell -d_0)  \Bigg\} \label{C1libre}\\
C_2&=-\frac{d_0^2}{K_a^2}\,\frac{H_f}{k_\mathrm{B}T}\,,  \label{C2libre}
\end{align}
where $H_f$ is the effective spring constant expressed in Eq.~\ref{ressortlibre}.

\subsection{Estimating $c'_0$}
\label{appcp0}
Let us start from the free energy per molecule in monolayer $+$ expressed in Eq.~\ref{fmolec}.
All the quantities involved in this expression are defined on the hydrophilic-hydrophobic interface $\mathcal{S}$ of the monolayer.

Let us consider a surface $\mathcal{S}'$ parallel to $\mathcal{S}$, and let us call $\delta$ the algebraic distance from $\mathcal{S}'$ to $\mathcal{S}$. To second order in the small dimensionless variables $c_1 \delta$ and $c_2 \delta$, where $c_1$ and $c_2$ denote the local principal curvatures of the monolayer (recall that $H=(c_1+c_2)/2$ and $K=c_1 c_2$), geometry gives~\cite{Safran}:
\begin{align}
\Sigma'&=\Sigma\left(1+2\,H\delta+K\delta^2\right)\,,\\
H'&=H+\left(K-2\,H^2\right)\delta\,,\\
K'&=K\,.
\end{align}
Hence, we can rewrite $f^+$ using variables defined on $\mathcal{S}'$, to second order:
\begin{align}
f^+&=\frac{1}{2}f''_0(\Sigma'-\Sigma_0)^2+ f_1\,H' + \left(f'_1-2\,f''_0\,\Sigma_0\,\delta\right)\left(\Sigma'-\Sigma_0\right)H'\nonumber\\
&+\left(f_2+2f''_0\,\Sigma_0^2\,\delta^2-2\,f'_1\,\Sigma_0\,\delta+2\,f_1\,\delta\right)H'^2+(f_K-f_1\,\delta)\,K'\nonumber\\
&+\alpha\,(\mathbf{\nabla}\Sigma')^2+\beta\,\nabla^2\Sigma'+\zeta\,(\nabla^2\Sigma')^2-\mu\,,
\end{align}
where we have neglected terms containing derivatives of order higher than two.

If $\mathcal{S}'$ is the neutral surface of the monolayer~\cite{Safran}, by definition, the curvature and the area variations are decoupled, which entails $f'_1=2\,f''_0\,\Sigma_0\,\xi$, where $\xi$ denotes the algebraic distance from the neutral surface to the hydrophilic-hydrophobic interface of the monolayer. Thus, given that $f''_0=K_a/(2\,\Sigma_0)$, $f_2=\kappa_0\,\Sigma_0$, and $f'_1/f_2=c'_0$ (see Methods, Sec.~\ref{Deriv}), we obtain 
\begin{equation}
c'_0\Sigma_0=\frac{K_a\,\xi}{\kappa_0}\,.
\label{c0primeexpr}
\end{equation}

\section*{Acknowledgments}
We thank Grace Brannigan for sharing with us some data corresponding to the simulations described in Ref.~\cite{Brannigan07}. We thank Mark Goulian for sharing with us a notebook containing some of the original calculations of Ref.~\cite{Goulian98}, and for email correspondence. We also acknowledge critical reading of our manuscript by Florent Bories.


\section*{Figure Legends}

\clearpage

\section*{Tables}

\begin{table}[htb]
\centering
\begin{tabular}{|l|l|l|l|l|}
\hline
&&Set 1&Set 2&Set 3\\
\hline
Free $s$&$H_f$ if $k'_a=0$&41&46&33\\
\hline
Free $s$&$k'_a$ if $H_f=H_\mathrm{exp}=115$&25&24&26\\
\hline
$s=0$&$H_0$ if $k'_a=0$&130&133&91\\
\hline
$s=0$&$k'_a$ if $H_0=H_\mathrm{exp}=115$&$<0$&$<0$&7.5\\
\hline
\end{tabular}
\caption{\textbf{Spring constant $H$ and constant $k'_a$ of monoolein.} The results are given both for the free-slope boundary condition (using Eq.~\ref{ressortlibre}) and for the zero-slope boundary condition $s=0$ (using Eq.~\ref{ressortfixe}). All values of $H$ and $k'_a$ are given in $\mathrm{mN/m}$. Negative values of $k'_a$ are not detailed since they would yield an instability for the monolayer Hamiltonian Eq.~\ref{onemore} in the present framework where $k''_a=0$. The different columns correspond to three different data sets for the parameters of the membrane. Set 1 corresponds to the data from~\cite{Chung94Nat} at $25^\circ\mathrm{C}$: $\kappa=3.6\times10^{-20}\,\mathrm{J}$, $c_0=-0.135\,\mathrm{nm}^{-1}$, $\bar\kappa= 8.8\times 10^{-22}\,\mathrm{J}$. Set 2 takes into account the corrections on $c_0$ and $\bar\kappa$ in~\cite{Templer95JPII}: $c_0=-0.196\,\mathrm{nm}^{-1}$, $\bar\kappa= -3.6\times 10^{-21}\,\mathrm{J}$. Set 3 corresponds to the data from~\cite{Vacklin00}: $\kappa=1.2\times10^{-20}\,\mathrm{J}$, $c_0=-0.503\,\mathrm{nm}^{-1}$, and $\bar\kappa= -1.2\times 10^{-21}\,\mathrm{J}$ deduced from $\bar\kappa/\kappa=-0.1$~\cite{Templer95JPII}. In all cases, we have taken $r_0=1\,\mathrm{nm}$~\cite{Lundbaek99}, $d_0= 2.46\,\mathrm{nm}$~\cite{Chung94BJ}, $\xi=-0.3\,\textrm{\AA}$~\cite{Templer95L}, $K_a=140\,\mathrm{mN/m}$~\cite{Hladky82, Lundbaek99}.}
\label{tableH}
\end{table}

\begin{table}[htb]
\centering
\begin{tabular}{|p{3.5cm}|p{1.2cm}|p{1.2cm}|p{1.2cm}|p{1.2cm}|p{1.2cm}|p{1.2cm}|}
\hline
Model&Ref.~\cite{Nielsen98}&Ref.~\cite{Nielsen98}&Ref.~\cite{Nielsen98}&Ours, with $k'_a=0$&Ours, with $k'_a=0$&Ours, with $k'_a=0$\\
\hline
Constants&Ref.~\cite{Goulian98}&Recent&Recent&Recent&Recent&Recent\\
\hline
Slope $s$&0.3&0.3&0&0.3&0&Free\\
\hline
$C_1\, \mathrm{[}10^{-3} \mathrm{(mN/m)}^{-1}\mathrm{]}$&354&282&480&292&502&339\\
\hline
$-C_2\, \mathrm{[}10^{-3} \mathrm{(mN/m)}^{-2}\mathrm{]}$&21.4&6.11&6.11&6.40&6.40&3.34\\
\hline
$\chi^2$&5.72&7.15&1.75&6.68&1.75&4.31\\
\hline
\end{tabular}
\caption{\textbf{Values of $C_1$ and $C_2$ obtained from the model of Ref.~\cite{Nielsen98} and from our model with $k'_a=0$.} The results are presented both for the fixed-slope boundary condition (see Eqs.~\ref{C1fixe} and~\ref{C2fixe}), with slopes $0$ and $0.3$, and for the free-slope boundary condition (see Eqs.~\ref{C1libre} and~\ref{C2libre}). The corresponding values of $\chi^2$ are also given. Recall that the best quadratic fit to the data of Ref.~\cite{Goulian98} yields $C_1 = 740\times 10^{-3} \,\mathrm{(mN/m)}^{-1}$ and $C_2 = -90.0\times 10^{-3}\,\mathrm{(mN/m)}^{-2}$ (see Fig.~\ref{fig:quadratic_fit}).}
\label{tableC1C2_1}
\end{table}

\begin{table}[htb]
\centering
\begin{tabular}{|l|l|l|l|}
\hline
Slope $s$&$0$&$0.3$&Free\\
\hline
$k'_a\, \mathrm{(mN/m)}$&$0$&45&30\\
\hline
$C_1\, \mathrm{[}10^{-3} \mathrm{(mN/m)}^{-1}\mathrm{]}$&502&490&490\\
\hline
$-C_2\, \mathrm{[}10^{-3} \mathrm{(mN/m)}^{-2}\mathrm{]}$&6.40&9.17&5.29\\
\hline
$\chi^2$&1.75&1.69&1.75\\
\hline
\end{tabular}
\caption{\textbf{Values of $k'_a$, $C_1$ and $C_2$ obtained from our model that yield the best agreement with the experimental results of Ref.~\cite{Goulian98}, analyzed with a quadratic fit (see Eq.~\ref{eq:quadratic} and Fig.~\ref{fig:quadratic_fit}).} Results are presented for the fixed-slope boundary condition (see Eqs.~\ref{C1fixe} and~\ref{C2fixe}), with slopes $0$ and $0.3$, and for the free-slope boundary condition (see Eqs.~\ref{C1libre} and~\ref{C2libre}).}
\label{tableC1C2_0}
\end{table}

\begin{table}[htb]
\centering
\begin{tabular}{|l|l|l|l|l|}
\hline
Slope $s$&0&0.3&-0.17&Free\\
\hline
$k'_a\, \mathrm{(mN/m)}$&23&78&0&60\\
\hline
$-C_2\, \mathrm{[}10^{-3} \mathrm{(mN/m)}^{-2}\mathrm{]}$&7.90&11.0&6.39&7.04\\
\hline
\end{tabular}
\caption{\textbf{Values of $k'_a$ and $C_2$ obtained from our model that yield the best agreement with the experimental results of Ref.~\cite{Goulian98} analyzed with the low-tension linear fit.} More precisely, these values of $k'_a$ and $C_2$ are associated with $C_1=0.62 \,\mathrm{(mN/m)}^{-1}$. Results are presented for the fixed-slope boundary condition (see Eqs.~\ref{C1fixe} and~\ref{C2fixe}), with slopes 0, $0.3$, $-0.17$, and for the free-slope boundary condition (see Eqs.~\ref{C1libre} and~\ref{C2libre}). }
\label{tableC1C2}
\end{table}


\begin{thebibliography}{10}
\providecommand{\url}[1]{\texttt{#1}}
\providecommand{\urlprefix}{URL }
\expandafter\ifx\csname urlstyle\endcsname\relax
  \providecommand{\doi}[1]{doi:\discretionary{}{}{}#1}\else
  \providecommand{\doi}{doi:\discretionary{}{}{}\begingroup
  \urlstyle{rm}\Url}\fi
\providecommand{\bibAnnoteFile}[1]{%
  \IfFileExists{#1}{\begin{quotation}\noindent\textsc{Key:} #1\\
  \textsc{Annotation:}\ \input{#1}\end{quotation}}{}}
\providecommand{\bibAnnote}[2]{%
  \begin{quotation}\noindent\textsc{Key:} #1\\
  \textsc{Annotation:}\ #2\end{quotation}}
\providecommand{\eprint}[2][]{\url{#2}}

\bibitem{Mouritsen_book}
Mouritsen OG (2005) Life -- as a matter of fat: the emerging science of
  lipidomics.
\newblock Springer, Berlin.
\input{Mouritsen_book}

\bibitem{Sackmann84}
Sackmann E (1984) Physical basis for trigger processes and membrane structures.
\newblock In: Chapman D, editor, Biological Membranes, London: Academic Press,
  volume~5. pp. 105-143.
\input{Sackmann84}

\bibitem{Helfrich73}
Helfrich W ({1973}) {Elastic properties of lipid bilayers - Theory and possible
  experiments}.
\newblock {Zeitschrift f\"ur Naturforschung C -- Journal of Biosciences} {28}:
  693-703.
\input{Helfrich73}

\bibitem{Owicki78}
Owicki JC, Springgate MW, McConnell H ({1978}) {Theoretical study of
  protein-lipid interactions in bilayer membranes}.
\newblock Proc Natl Acad Sci USA {75}: 1616-1619.
\input{Owicki78}

\bibitem{Owicki79}
Owicki JC, McConnell H ({1979}) {Theory of protein-lipid and protein-protein
  interactions in bilayer membranes}.
\newblock Proc Natl Acad Sci USA {76}: 4750-4754.
\input{Owicki79}

\bibitem{Mouritsen84}
Mouritsen OG, Bloom M ({1984}) {Mattress model of lipid-protein interactions in
  membranes}.
\newblock Biophys J {46}: 141-153.
\input{Mouritsen84}

\bibitem{Huang86}
Huang HW ({1986}) {Deformation free energy of bilayer membrane and its effect
  on gramicidin channel lifetime}.
\newblock Biophys J {50}: {1061-1070}.
\input{Huang86}

\bibitem{Kelkar07}
Kelkar DA, Chattopadhyay A (2007) The gramicidin ion channel: A model membrane
  protein.
\newblock Biochim Biophys Acta -- Biomembr 1768: 2011-2025.
\input{Kelkar07}

\bibitem{OConnell90}
{O'Connell} AM, {Koeppe II} RE, Andersen OS (1990) Kinetics of gramicidin
  channel formation in lipid bilayers: transmembrane monomer association.
\newblock Science 250: 1256-1259.
\input{OConnell90}

\bibitem{Lundbaek10}
Lundbaek JA, {Koeppe II} RE, Andersen OS ({2010}) {Amphiphile regulation of ion
  channel function by changes in the bilayer spring constant}.
\newblock Proc Natl Acad Sci USA {107}: {15427-15430}.
\input{Lundbaek10}

\bibitem{Helfrich90}
Helfrich P, Jakobsson E (1990) Calculation of deformation energies and
  conformations in lipid membranes containing gramicidin channels.
\newblock Biophys J 57: 1075-1084.
\input{Helfrich90}

\bibitem{Dan93}
Dan N, Pincus P, Safran SA ({1993}) {Membrane-induced interactions between
  inclusions}.
\newblock {Langmuir} {9}: {2768-2771}.
\input{Dan93}

\bibitem{ArandaEspinoza96}
Aranda-Espinoza H, Berman A, Dan N, Pincus P, Safran S ({1996}) {Interaction
  between inclusions embedded in membranes}.
\newblock Biophys J {71}: 648.
\input{ArandaEspinoza96}

\bibitem{Brannigan06}
Brannigan G, Brown FLH ({2006}) {A consistent model for thermal fluctuations
  and protein-induced deformations in lipid bilayers}.
\newblock Biophys J {90}: {1501-1520}.
\input{Brannigan06}

\bibitem{Brannigan07}
Brannigan G, Brown FLH ({2007}) Contributions of gaussian curvature and
  nonconstant lipid volume to protein deformation of lipid bilayers.
\newblock Biophys J {92}: {864-876}.
\input{Brannigan07}

\bibitem{West09}
West B, Brown FLH, Schmid F ({2009}) Membrane-protein interactions in a generic
  coarse-grained model for lipid bilayers.
\newblock Biophys J {96}: {101-115}.
\input{West09}

\bibitem{Elliott83}
Elliott JR, Needham D, Dilger JP, Haydon DA (1983) The effects of bilayer
  thickness and tension on gramicidin single-channel lifetime.
\newblock Biochim Biophys Acta -- Biomembr 735: 95-103.
\input{Elliott83}

\bibitem{Goulian98}
Goulian M, Mesquita ON, Fygenson DK, Nielsen C, Andersen OS, et~al. ({1998})
  {Gramicidin channel kinetics under tension}.
\newblock Biophys J {74}: {328-337}.
\input{Goulian98}

\bibitem{Safran}
Safran SA (1994) Statistical thermodynamics of surfaces, interfaces and
  membranes.
\newblock Addison-Wesley.
\input{Safran}

\bibitem{Bitbol11_stress}
Bitbol AF, Peliti L, Fournier JB ({2011}) {Membrane stress tensor in the
  presence of lipid density and composition inhomogeneities}.
\newblock Eur Phys J E {34}: {53}.
\input{Bitbol11_stress}

\bibitem{Fournier99}
Fournier JB ({1999}) {Microscopic membrane elasticity and interactions among
  membrane inclusions: interplay between the shape, dilation, tilt and
  tilt-difference modes}.
\newblock Eur Phys J B {11}: {261-272}.
\input{Fournier99}

\bibitem{Nielsen98}
Nielsen C, Goulian M, Andersen OS ({1998}) {Energetics of inclusion-induced
  bilayer deformations}.
\newblock Biophys J {74}: {1966-1983}.
\input{Nielsen98}

\bibitem{Watson11}
Watson MC, Penev ES, Welch PM, Brown FLH ({2011}) {Thermal fluctuations in
  shape, thickness, and molecular orientation in lipid bilayers}.
\newblock J Chem Phys {135}: {244701}.
\input{Watson11}

\bibitem{Israelachvili}
Israelachvili JN (1992) Intermolecular and surface forces, Second edition.
\newblock Academic Press.
\input{Israelachvili}

\bibitem{Sharp91}
Sharp KA, Nicholls A, Fine RF, Honig B ({1991}) {Reconciling the magnitude of
  the microscopic and macroscopic hydrophobic effects}.
\newblock {Science} {252}: 106-109.
\input{Sharp91}

\bibitem{May99}
May S, Ben-Shaul A ({1999}) {Molecular theory of lipid-protein interaction and
  the L$_\alpha$-H$_{||}$ transition}.
\newblock Biophys J {76}: {751-767}.
\input{May99}

\bibitem{Hladky82}
Hladky SB, Gruen DWR ({1982}) {Thickness fluctuations in black lipid
  membranes}.
\newblock Biophys J {38}: 251-258.
\input{Hladky82}

\bibitem{Rawicz00}
Rawicz W, Olbrich KC, McIntosh T, Needham D, Evans E (2000) Effect of chain
  length and unsaturation on elasticity of lipid bilayers.
\newblock Biophys J 79: 328-339.
\input{Rawicz00}

\bibitem{May07}
May ER, Narang A, Kopelevich DI ({2007}) {Role of molecular tilt in thermal
  fluctuations of lipid membranes}.
\newblock Phys Rev E {76}: {021913}.
\input{May07}

\bibitem{May07b}
May ER, Narang A, Kopelevich DI ({2007}) {Molecular modeling of key elastic
  properties for inhomogeneous lipid bilayers}.
\newblock {Mol Simulat} {33}: {787-797}.
\input{May07b}

\bibitem{Watson12}
Watson MC, Brandt EG, Welch PM, Brown FLH ({2012}) {Determining biomembrane
  bending rigidities from simulations of modest size}.
\newblock Phys Rev Lett {109}: {028102}.
\input{Watson12}

\bibitem{Lindahl00}
Lindahl E, Edholm O ({2000}) {Mesoscopic undulations and thickness fluctuations
  in lipid bilayers from molecular dynamics simulations}.
\newblock Biophys J {79}: {426-433}.
\input{Lindahl00}

\bibitem{Marrink01}
Marrink SJ, Mark AE ({2001}) {Effect of undulations on surface tension in
  simulated bilayers}.
\newblock J Phys Chem B {105}: {6122-6127}.
\input{Marrink01}

\bibitem{Lundbaek99}
Lundbaek JA, Andersen OS ({1999}) {Spring constants for channel-induced lipid
  bilayer deformations estimates using gramicidin channels}.
\newblock Biophys J {76}: {889-895}.
\input{Lundbaek99}

\bibitem{Kim12}
Kim T, Lee KI, Morris P, Pastor RW, Andersen OS, et~al. ({2012}) {Influence of
  hydrophobic mismatch on structures and dynamics of gramicidin A and lipid
  bilayers}.
\newblock Biophys J {102}: 1551-1560.
\input{Kim12}

\bibitem{Kolb77}
Kolb HA, Bamberg E (1977) Influence of membrane thickness and ion concentration
  on the properties of the gramicidin {A} channel: autocorrelation, spectral
  power density, relaxation and single-channel studies.
\newblock Biochim Biophys Acta -- Biomembr 464: 127-141.
\input{Kolb77}

\bibitem{Hwang03}
Hwang TC, {Koeppe II} RE, Andersen OS (2003) Genistein can modulate channel
  function by a phosphorylation-independent mechanism: importance of
  hydrophobic mismatch and bilayer mechanics.
\newblock Biochemistry 42: 13646-13658.
\input{Hwang03}

\bibitem{Chung94Nat}
Chung H, Caffrey M (1994) {The curvature elastic-energy function of the
  lipid-water cubic mesophase}.
\newblock {Nature} 368: 224-226.
\input{Chung94Nat}

\bibitem{Chung94BJ}
Chung H, Caffrey M ({1994}) {The neutral area surface of the cubic mesophase:
  location and properties}.
\newblock Biophys J {66}: {377-381}.
\input{Chung94BJ}

\bibitem{Templer95L}
Templer RH ({1995}) {On the area neutral surface of inverse bicontinuous cubic
  phases of lyotropic liquid crystals}.
\newblock {Langmuir} {11}: {334-340}.
\input{Templer95L}

\bibitem{Templer95JPII}
Templer RH, Turner DC, Harper P, Seddon JM ({1995}) {Corrections to some models
  of the curvature elastic energy of inverse bicontinuous cubic phases}.
\newblock {J Phys II France} {5}: {1053-1065}.
\input{Templer95JPII}

\bibitem{Vacklin00}
Vacklin H, Khoo BJ, Madan KH, Seddon JM, Templer RH ({2000}) {The bending
  elasticity of 1-monoolein upon relief of packing stress}.
\newblock {Langmuir} {16}: {4741-4748}.
\input{Vacklin00}

\bibitem{Szule02}
Szule JA, Fuller NL, Rand RP ({2002}) {The effects of acyl chain length and
  saturation of diacylglycerols and phosphatidylcholines on membrane monolayer
  curvature}.
\newblock Biophys J {83}: {977-984}.
\input{Szule02}

\bibitem{Leikin96}
Leikin S, Kozlov MM, Fuller NL, Rand RP ({1996}) {Measured effects of
  diacylglycerol on structural and elastic properties of phospholipid
  membranes}.
\newblock Biophys J {71}: {2623-2632}.
\input{Leikin96}

\end{thebibliography}
\end{document}